\documentclass[a4paper,12pt]{article}
\usepackage{amssymb, amsmath}
\usepackage[english]{babel}
\usepackage[applemac]{inputenc}
\usepackage[dvips]{graphicx}
\usepackage{amsfonts}
\usepackage{mathrsfs}
\usepackage{color}
\newcommand{\ud}{\mathrm{d}}
\newcommand{\shalf}{{\hbox{\small{$\frac{1}{2}\hskip 0.01cm$}}}}
\newtheorem{theorem}{Theorem}[section]
\newtheorem{proposition}{Proposition}[section]
\newtheorem{lemma}{Lemma}[section]
\newtheorem{remark}{Remark}[section]

\newtheorem{definition}{Definition}[section]

\begin{document}

\begin{center}

{\Large{\textbf{Exactness of the Bogoliubov approximation\\ in random external potentials}}}

\vspace{1cm}

\textbf{Thomas Jaeck\footnote{PhD student at UCD (Dublin, Ireland) and Universit\'{e} de la M\'{e}diterran\'{e}e
(Aix-Marseille II, France), \\
e-mail: Thomas.Jaeck@ucdconnect.ie, phone: +353 1 7162571}}\\
School of Mathematical Sciences, University College Dublin\\
Belfield, Dublin 4, Ireland

\vspace{1cm}

\textbf{Valentin A. Zagrebnov\footnote{e-mail: Valentin.Zagrebnov@cpt.univ-mrs.fr, phone: +33 491 26 95 04}}\\
Universit\'{e} de la M\'{e}diterran\'{e}e (Aix-Marseille II), \\ Centre de Physique Th\'{e}orique - UMR 6207,
Luminy - Case 907 \\ 13288 Marseille, Cedex 09, France

\vspace{1cm}

\begin{abstract}
\noindent We investigate the validity of the Bogoliubov $c$-number approximation in the case of interacting Bose-gas
in a \textit{homogeneous random} media. To take into account the possible occurence of \textit{type} III generalized
Bose-Einstein condensation (i.e. the occurrence of condensation in an infinitesimal band of low kinetic energy
modes \emph{without} macroscopic occupation of any of them) we generalize the $c$-number substitution procedure
to this band of modes with low momentum. We show that, as in the case of the one-mode condensation for
translation-invariant interacting systems, this procedure has \textit{no effect} on the exact value of the 
pressure in the
thermodynamic limit, assuming that the $c$-numbers are chosen according to a suitable \textit{variational principle}.
We then discuss the relation between these $c$-numbers and the (total) density of the condensate.
\end{abstract}

\end{center}

\noindent\textbf{Keywords:} Generalized Bose-Einstein Condensation, Random
Potentials, Bogoliubov\\ $c$-number Approximation, Bogoliubov Quasi-Averages, Berezin-Lieb Inequalities.

\vspace{0.5cm}

\noindent \textbf{PACS:} 05.30.Jp, 03.75.Fi, 67.40.-w  \\
\textbf{AMS:} 82B10, 82B23, 81V70

\newpage

\section{Motivation}
\setcounter{equation}{0}
\renewcommand{\theequation}{\arabic{section}.\arabic{equation}}
In 1947 Bogoliubov \cite{B} proposed an \textit{ansatz} that for large Boson systems the particle creation and
annihilation operators: $a^*_0$, $a_0$, corresponding to the zero momentum $k=0$, can be replaced by complex numbers.
This procedure is known as the Bogoliubov $c$-number approximation. It is based on the idea that these creation and
annihilation operators, when divided by the square root of the volume, $V$, of the region $\Lambda$ containing 
the system,
can be expressed as space averages:
\begin{eqnarray*}
\frac{a^{\#}_0}{\sqrt{V}} \, := \frac{1}{V} \int_{\Lambda} \ud x \, a^{\#}(x) ,
\end{eqnarray*}
where $a^{\#}(x)$ are the usual local creation and annihilation operators. Therefore for translation-invariant ergodic 
states these operators should converge to multiples of the identity:
\begin{eqnarray*}
\frac{a^{\#}_{0}}{\sqrt{V}} \, \rightarrow \, \alpha^{\#} \ ,
\end{eqnarray*}
in some \textit{weak sense}, see e.g. \cite{FPV}. These ideas were exploited to construct a various truncations of the
full interacting boson Hamiltonian. We refer the reader to, e.g., \cite{ZB} for a review of these models and to 
\cite{PZ}, \cite{BZ} for some recent applications.

The most spectacular result derived from this ansatz was its application to a homogeneous model of a weakly interacting Bose gas
\cite{B}, \cite{ZB}, which provided explicitly a spectrum of collective excitations satisfying the Landau criteria of
\textit{superfluidity}. Note that this \textit{microscopic} theory of superfluidity is also based on two other
Bogoluibov's \textit{ansatse}: the occurrence in the weakly interacting boson system of condensation in the $k=0$ mode
and the truncation of the full Hamiltonian, keeping only the \lq\lq dominant" terms, that is
those that involves at least two particles from the condensate.

Recently, see \cite{HM}, \cite{KT}, the Bogoliubov approximation has been used to study interacting bosons systems in
homogeneous external random potentials, where the notion of the ground-state as well as the existence
of the condensation are quite subtle. The aim of our paper is to study the validity of the Bogoliubov scheme for this
kind of models.

The first rigorous result concerning the Bogoliubov $c$-number approximation was due to Ginibre \cite{G}. For a
homogeneous boson gas with a two-body superstable interaction he proved that the Bogoliubov \textit{ansatz}, 
supplemented by a self-consistency \textit{condensate equation} which maximizes the \textit{approximated pressure} 
with respect to the c-number $\alpha$, gives the right pressure in the thermodynamic limit.
A transparent and elegant proof of this and other related results has been recently given by Lieb et
al \cite{LSY}, using, in particular, the Berezin-Lieb and the Bogoliubov convexity inequalities.
The paper \cite{LSY} also investigates a delicate point: namely, whether the value of the variational parameter
$\alpha_{\max}$ maximizing the approximating pressure \textit{coincides} with the \textit{total} condensate density.
There it was shown that the maximizer $\alpha_{\max}$ corresponds to the zero-mode condensate
density \emph{if} the gauge symmetry
breaking term (quasi-average \textit{sources}) of the form: $\sqrt{V} (\eta \, a^{*}_{0} + \overline{\eta} \, a_{0})$
is added to the full Hamiltonian. The idea of breaking the gauge-symmetry of the quantum Gibbs state for $k=0$
is due to Bogoliubov, \cite{Bog}.
This forces the \textit{totality} of the condensate to be concentrated in the
zero-mode (ground state). It must be switched off ($\eta = |\eta|\, e^{i \arg \eta}$, $|\eta|\rightarrow 0$ with a fixed
gauge $\phi:=\arg \eta$) \textit{after} the thermodynamic limit to produce a limiting Gibbs state. The 
expectation defined by this state is called
the Bogoliubov \textit{quasi-average} with respect to this source. The quasi-averages of the operators 
$a^{\#}_{0}/\sqrt{V}$
coincide with $|\alpha_{\max}|\, e^{\pm i \phi}$, i.e., $|\alpha_{\max}|^2$ is equal to the \textit{total} condensate
density \cite{LSY}. It has been argued in \cite{Bog}, \cite{LSY} that this quasi-average is
the only physically reliable quantity to describe Bose Einstein condensation.

We emphasize this point, because it has been known since \cite{vdBLP} that the Bose-Einstein condensation in the
\textit{gauge-invariant} systems does \emph{not} necessarily imply a macroscopic occupation of the ground state only.
Indeed, the condensate can be spread over many (possibly \textit{infinitely} many) quantum states, and in some cases,
\emph{none} of these states are macroscopically occupied. In all cases however, the \textit{total} amount of condensate
in an arbitrary small band of energy in the vicinity of the ground-state is the same, a phenomenon known as
\textit{generalized} condensation in the terminology of van den Berg-Lewis-Pul\'{e} \cite{vdBLP}. 

After Ginibre \cite{G} it was tempting to conjecture that the limiting value of the solution to the finite volume
\textit{condensate equation} yields the correct condensate density under all circumstances. By means of a
counter example it was shown in \cite{BdSP} that this is not so.
Although the Bogoliubov $c$-number approximation still gives the right pressure in these systems, it has been shown
for the mean-field boson gas \cite{BdSP}, that the solution $\alpha_{\max}$ of the condensate equation does 
\textit{not} provide the ground-state condensate density $\rho_0$, but the \textit{total} amount of the generalized 
condensate, i.e., $\rho_0 \leq |\alpha_{\max}|^2$.
This was in a striking contrast with a general conviction that the \textit{gauge-invariant}, translation invariant
boson systems always manifest the \textit{total} amount condensation in the ground state, and hence that the 
Bogoliubov $c$-number approximation in the zero-mode coincides with this amount.

We return to this point at the end of the present paper where we discuss our main result (Theorem
\ref{convergence-approximated-pressure}) in relation to the results of \cite{G}, \cite{BdSP}, \cite{Bog}, and \cite{LSY}.

The aim of our paper is to prove that the Bogoliubov $c$-number approximation to
interacting Bose-gas can be extended to the case of homogeneous
random external potentials (random media). Note that the arguments in
\cite{LSY} are valid for inhomogeneous systems and allows to treat condensation in many modes, as long as 
its number is much smaller than the volume $V$. In the case of random media this problem is more complicated.
First, we have to investigate a random inhomogeneous system, albeit with non-random properties 
(\textit{self-averaging}) in the thermodynamic limit  \cite{LGP}. Secondly, since the randomness seems to force the 
generalised condensation to be of type III, as we disucussed in \cite{JPZ1}, the number of modes occupied by 
the condensation in the random potentials is \textit{a priori} of the order of $V$. 

Recall that for non-interacting (perfect) bosons systems embedded into a \textit{bona fide} random potential, 
the generalized condensation occurs even in low-dimensional systems (see \cite{KL1, KL2, LS} and \cite{LPZ}, 
\cite{JPZ1} for a review) even though it does not occur for the corresponding translation-invariant systems. 
This is caused by the fact that the one-particle density of states has \emph{Lifshitz tails}, that is 
an extremely low density of quantum states near the bottom of the 
spectrum, a well-known feature of random systems widely believed
to be associated with the existence of localised eigenstates. Kac and 
Luttinger \cite{KL1, KL2} conjectured that in a homogeneous random potential condensation occurs in the state 
with the lowest energy (ground state). Indeed, one can check this conjecture \cite{LZ} for the particular case of the 
Luttinger-Sy model \cite{LS}. Recently in \cite{JPZ1}, it was shown that whenever condensation occurs in the random 
perfect Bose-gas, then there is a \textit{generalized} condensation in the kinetic-energy momentum states, and the 
both densities have the same value. This result can be partially extended to some
simple models of interacting Bose gases (mean-field models). It was proved in \cite{JPZ2} that under a fairly weak
assumption about localization this condensation is of the \textit{type} III, i.e. there is no macroscopic accumulation
of particles in any single momentum state. We conjecture that this holds also for general interacting systems in the
presence of homogeneous random potentials.

For the above reasons, we would like to follow the general philosophy of the Bogoliubov
$c$-number substitution ansatz and extend it to a \textit{generalized} Bogoliubov 
approximation, in order to cover the possible case of \textit{type} III generalized condensation.
By this, we mean a replacement of all creation/annihilation operators corresponding to momentum 
states with kinetic energy $\varepsilon_{k}$ in the energy band $0 \leq\varepsilon_{k} < \delta$ by complex numbers
$\{\sqrt{V}\alpha^{\sharp}_{k}\}_{\{k: \varepsilon_{k} < \delta\}}$.\\
First we show that this extension of the  Bogoliubov approximation applied to interacting Bose-gas in homogeneous
random potentials is valid as far as the pressure is concerned (Section \ref{Section-convergence-bogoliubov-pressure}). 
In this case as in the case one-mode condensation the corresponding
trial pressure is maximized with respect to these complex numbers and then one lets the parameter $\delta \rightarrow 0$
\textit{after} the thermodynamic limit. Note that for each realization of the random potential, the system is not
translation invariant and the minimizer has a random value \textit{before} the thermodynamic limit.
For this reason, the 
proof use the stationarity and ergodicity of the random potential with respect to the space translations.\\
Finally, we discuss the variational problem established for the pressure (Section \ref{Section-condensation}). 
In particular, we highlight the fact that the link between the $c$-numbers that maximize the trial pressure and 
the structure of the condensate is highly non-trivial. By the mean of a simple example, we show that the Bogoliubov
quasi-average technique of adding external sources \cite{Bog} is not \lq\lq satisfactory" in this case, since we suspect 
that the generalised condensate should be of type III, while this procedure is able to drastically 
alter the fine structure of the generalised condensate. This brings us back to the discussion of the gauge symmetry 
breaking and the physical reliability of the Bogoliubov quasi-averages in the case of condensation \cite{Bog}, 
\cite{LSY}.

\section{Model and definitions}\label{Section-models-definitions}
\setcounter{equation}{0}
\renewcommand{\theequation}{\arabic{section}.\arabic{equation}}
Let $\{\Lambda_{l}:= (-l/2, l/2)^{d}\}_{l \geqslant 1}$ be a sequence of hypercubes of side $l$ in
$\mathbb{R}^{d}, d \geqslant 1$, centered at the origin of coordinates with volumes $V_{l} = l^{d}$.
We consider a system of identical bosons, of mass $m$, contained
in $\Lambda_{l}$. For simplicity, we use a system of units such that $\hbar = m = 1$. First we define the
self-adjoint one-particle kinetic-energy operator of our system by:
\begin{eqnarray}\label{Kinetic-energy-operator}
h_{l}^{0} := - \, \shalf \ \Delta_P,
\end{eqnarray}
acting {in} the Hilbert space $\mathscr{H}_{l} := L^{2} (\Lambda_{l})$. The subscript $P$ stands for the
\textit{periodic} boundary condition. We denote by $\{\psi_{k}^{l}, \varepsilon_{k}^{l}\}_{k \in \Lambda^{*}}$ the
set of normalized eigenfunctions and eigenvalues corresponding to operator $h_{l}^{0}$
\begin{eqnarray}\label{kinetic-functions}
\psi_{k}^{l}(x) \, = \, \frac{1}{\sqrt{V_{l}}} \ e^{i k \cdot x}, \quad \varepsilon_{k}^{l} \, = \,
\shalf \ k^{2} \ ,
\end{eqnarray}
and $\Lambda^{*}_{l}$ is the usual dual space to $\Lambda_{l}$, that is $\Lambda^{*}_{l}:=
\{k \in \mathbb{R}^{d}: k^{2} = {(2\pi n)^{2}}/{l^{2}}, n \in \mathbb{Z}^{d}\}$.
Finally, we denote by $\nu_{l}^{0}$ the finite-volume \textit{integrated density of states} (IDS), that is,
\begin{eqnarray}\label{density-states-kinetic-energy-operator}
\nu_{l}^{0} (E) \, := \, \frac{1}{V_{l}} \, \sharp \{k \in \Lambda^{*}_{l}: \varepsilon_{k}^{l} \leqslant E\},
\end{eqnarray}
and we let $\nu^{0} (E) := \lim_{l \rightarrow \infty} \nu_{l}^{0} (E)$. Note that the limiting IDS, $\nu^{0} (E)$,
has support on $[0,\infty)$ and that it is known explicitly: $\nu^{0} (E) = C_d \, E^{d/2}$ 
(the \textit{Weyl formula}).
\begin{definition}\label{rand-potent}
We define an external random potential $v^{(\cdot)}(\cdot)\, : \,\Omega \times \mathbb{R}^{d}
\rightarrow \mathbb{R}, \ x \mapsto v^{\omega}(x)$ as a measurable random field on a probability space
$(\Omega, \mathcal{F}, \mathbb{P})$, satisfying the following conditions:\\
{\rm{(i)}} $v^{\omega}, \omega\in \Omega$, is non-negative;\\
{\rm{(ii)}} $p := \mathbb{P} \{\omega: v^{\omega}(0) = 0\} < 1$.\\
As usual, we assume also {\rm{(}}see e.g. Appendix B in {\rm{\cite{JPZ1}}}{\rm{)}} that this field  is: \\
{\rm{(iii)}} homogeneous (stationary) and ergodic with respect to the group $\{\tau_x\}_{x\in \mathbb{R}^{d}}$
of probability preserving translations on $(\Omega, \mathcal{F}, \mathbb{P})$ ;\\
{\rm{(iv)}} $\varphi$-mixing for $\Sigma_{\Lambda}$-measurable functions, where $\Sigma_{\Lambda}$ is
the $\sigma$-algebra
generated by the field $\{v^{\omega}(x)\}_{x\in \Lambda}$ for $\Lambda \subset \mathbb{R}^{d}$.

\end{definition}
Then the corresponding self-adjoint random Schr\"{o}dinger operator acting in $\mathscr{H} := L^{2} (\mathbb{R}^d)$ 
is a perturbation of the kinetic-energy operator:
\begin{eqnarray}\label{Schrodinger-operator-inf}
h^{\omega} := - \shalf \ \Delta \, \dotplus \, v^{\omega} ,
\end{eqnarray}
defined as a sum in the \textit{quadratic-forms} sense. The restriction to the box $\Lambda_{l}$,
is specified by the periodic boundary conditions and for regular potentials one gets the self-adjoint
operator:
\begin{eqnarray}\label{Schrodinger-operator}
h_{l}^{\omega} := \left( - \shalf \ \Delta \, + \, v^{\omega} \right)_{P} = h_{l}^{0} \dotplus \, v_{l}^{\omega},
\end{eqnarray}
acting in $ \mathscr{H}_l$, where $v_{l}^{\omega}$ is the restriction of $v^{\omega}$ to $\Lambda_{l}$.
We denote by $\{\phi_{i}^{\omega,l}, E_{i}^{\omega,l}\}_{i \geqslant 1}$ the
set of normalized eigenfunctions and the corresponding eigenvalues of the random operator $h^{\omega}_{l}$.
We order the eigenvalues
(counting the multiplicity) in such a way that $E_{1}^{\omega,l} \leqslant E_{2}^{\omega,l} \leqslant E_{3}^{\omega,l}
\dots \,\,$.\\
Note that the \textit{non-negativity} of the random potential implies {that} $E_{1}^{\omega,l} > 0$.
So, for convenience we \textit{assume} also that in the thermodynamic limit  the lowest edge of this random
one-particle spectrum $\sigma(h^{\omega}_{l})$ satisfies the fifth condition:\\
${\rm{(v)}} \lim_{l \rightarrow \infty} E_{1}^{\omega,l} = 0 \ ,$ \emph{almost surely} (a.s.)
with respect to the probability $\mathbb{P}$.
\begin{remark}\label{rand-potent=0}
Note that {\rm{(v)}} is in fact an implicit condition on the random potential saying that a.s.,
one can find a sequence of regions \emph{(gaps)} with $v^{\omega}(x) = 0$, with volumes tending to infinity
in the van Hove sense.
\end{remark}
Now, we turn to the many-body problem. Let $\mathscr{F}_{l}:= \mathscr{F}_{l}(\mathscr{H}_{l})$ be the
symmetric Fock space constructed over $ \mathscr{H}_{l}$. Then $H_{l}^{0}:={\rm{d\Gamma}}(h_l^\omega)$ denotes
the second quantization of the \textit{one-particle} Schr\"{o}dinger operator $h_{l}^{\omega}$ in $\mathscr{F}_{l}$.
For simplicity, we omit the explicit mention of the randomness of the Hamiltonians and all related quantities,
unless this is necessary for the sake of clarity.
Since for any $\omega\in\Omega$ the one-particle eigenstates $\{\phi_{i}:= \phi_{i}^{\omega,l}\}_{i \,\geq 1}$ of
$h_{l}^{\omega}$ form a basis in $\mathscr{H}_{l}$, the operator $H_{l}^{0}$ acting in $\mathscr{F}_{l}$ can be
expressed as:
\begin{equation}\label{multi-part-perfect-Hamiltonian}
H_{l}^{0} =  \sum_{i \geqslant 1}  E_{i}^{\omega,l} \   a^{*}(\phi_{i}) a(\phi_{i})
 \, = \, \sum_{i \geqslant 1}  E_{i}^{\omega,l} N_{l}(\phi_{i}) \ .
\end{equation}
Here $a^{*}(\phi_{i}),  a(\phi_{i})$ are the creation and annihilation operators, satisfying the boson
\textit{Canonical Commutation Relations}, and $N_{l}(\phi_{i})$ is the particle-number operator
in the state $\phi_{i}^{\omega,l}$. Note that, since $[h_{l}^{0},h_{l}^{\omega}] \neq 0$,
the Hamiltonian (\ref{multi-part-perfect-Hamiltonian}) cannot be expressed as a function of the operators
$N_{l}(\psi_{k}^{l})$ in the kinetic-energy eigenstates (\ref{kinetic-functions}).

\noindent (vi) Below we assume that the particles interact through a suitable \textit{non-negative} two-body
\textit{translation-invariant} potential $\Phi(x,y) := u(|x-y|)$. More precisely, we assume that
the function $u$ has a continuous, bounded Fourier transformation $ \hat{u}(q)$, and there is $\gamma < \infty$
such that $\vert \hat{u}(q) \vert < \gamma$ for all $q \in \Lambda^{*}_{l}$ and for all $l$.
For example, one can choose ${u}\in L^1(\mathbb{R}^d)$.

Then the second quantization of interaction: $U_l:={\rm{d\Gamma}}(\Phi)$ has a simple form in
the translation-invariant basis $\{\psi_{k}:=\psi_{k}^{l}\}_{k\in\Lambda^{*}_{l}}$ of the kinetic-energy operator
$h_{l}^{0}$:
\begin{equation}\label{U}
U_l =  \frac{1}{2V_{l}} \sum_{q,k,k' \in \Lambda_{l}^{*}} \
\hat{u}(q) \ a^{*}(\psi_{k+q}) a^{*}(\psi_{k'-q}) a(\psi_{k'}) a(\psi_{k}) \ .
\end{equation}
The full Hamiltonian with the chemical potential $\mu$ included has the form:
\begin{eqnarray}\label{multi-part-interacting-Hamiltonian}
H_{l} (\mu) \, := \, H_{l}^{0} - \mu N_{l} + U_l
\end{eqnarray}
Note that the creation and annihilation operators in the interaction term (\ref{U}) are in the momentum eigenstates
$\{\psi_{k}\}_{k\in\Lambda^{*}_{l}}$, although the perfect Bose-gas Hamiltonian (\ref{multi-part-perfect-Hamiltonian})
is not diagonal when expressed in this basis.

By $\langle - \rangle_{l}$ we denote below the grand-canonical equilibrium state defined by the Hamiltonian 
$H_{l}(\mu)$:
\begin{eqnarray*}
\langle A\rangle_{l}(\beta,\mu) :=
\frac{1}{\Xi_{l}(\beta, \mu)} \, {\rm{Tr}}_{ \mathscr{F}_{l}}
\exp ( -\beta H_{l}(\mu) ),
\end{eqnarray*}
and by $p_{l}(\beta,\mu)$ its associated grand-canonical pressure
\begin{eqnarray}\label{press}
p_{l}(\beta,\mu) \, := \, \frac{1}{\beta V_{l}} \ln \Xi_{l}(\beta, \mu),
\end{eqnarray}
where
\begin{eqnarray*}
\Xi_{l}(\beta, \mu) \, := \, {\rm{Tr}}_{ \mathscr{F}_{l}}
\exp ( -\beta H_{l}(\mu) )
\end{eqnarray*}
is the corresponding grand-canonical partition function.

It is known that the pressure of the corresponding non-random model (i.e. for $v^{\omega}(x) = 0$) with a
\textit{bona fide} interaction exists and is independent of the boundaries condition for a large class of them,
including the periodic case, see e.g. \cite{Rob}. The proof of this statement consists essentially in showing
the existence of the Dirichlet pressure using sub-additivity
\begin{eqnarray*}
p_{\Lambda}^{D}(\beta,\mu) \, \geqslant \, p_{\Lambda'}^{D}(\beta,\mu) + p_{\tau_{x}\Lambda''}^{D}(\beta,\mu)
\end{eqnarray*}
where $\Lambda', \Lambda''$ are disjoints subsets of $\Lambda$, and $\tau_{x}$ denotes translation by $x$.
The exact value of $x$ is chosen according to the usual \textit{tempering condition} required of the two-body
interaction potential (vi). Then, using translation invariance of the non-random model, one obtains
\begin{eqnarray}\label{Dirichlet-pressure-sub-additivity-non-random-case}
p_{\Lambda}^{D}(\beta,\mu) \, \geqslant \, p_{\Lambda'}^{D}(\beta,\mu) + p_{\Lambda''}^{D}(\beta,\mu) \ ,
\end{eqnarray}
and the boundeness of the pressure, which is provided by the \textit{superstability} of the interaction
($u\geq 0$, (vi)), leads to the existence and finiteness of the limiting pressure for any $\mu$. Then, one can show using
functional integration techniques, see \cite{AN}, that the others boundary conditions converge to the same limit.

The last part of this prove can be carried through verbatim in the presence of the external random potential. However,
because of the lack of translation invariance in the random case, the inequality
(\ref{Dirichlet-pressure-sub-additivity-non-random-case}) for the Dirichlet pressure is modified as follows:
\begin{eqnarray}\label{Dirichlet-pressure-sub-additivity-random-case}
p_{\Lambda}^{D, \omega}(\beta,\mu) \, \geqslant \, p_{\Lambda'}^{D, \omega}(\beta,\mu) +
p_{\tau_{x}\Lambda''}^{D,\omega}(\beta,\mu)
\, = \, p_{\Lambda'}^{D, \omega}(\beta,\mu) + p_{\Lambda''}^{D,\tau_{x}\omega}(\beta,\mu) \ .
\end{eqnarray}
We used the \textit{stationarity} of the random potential in the last identity. To prove the the existence of the
thermodynamic limit one can use the Kingman \textit{sub-additive} ergodic theorem, see \cite{Ste}:
\begin{proposition}\label{Kingman}
Let $\tau$ be measure preserving transformation of the probability space $(\Omega, \mathcal{F},\mathbb{P})$
and $\{g_n\}_{n\geq 1}$ be a sequence of functions $g_n \in L^1(\Omega, \mathcal{F},\mathbb{P})$ satisfying the 
condition:
\begin{equation}\label{g}
g_{n+m}(\omega) \leq g_{n}(\omega) + g_{m}(\tau^{n} \omega) \
\end{equation}
Then one gets that
\begin{equation}\label{lim-g}
a.s.-\lim_{n\rightarrow\infty} g_{n}(\omega)/n = g(\omega) \ ,
\end{equation}
where the function $g(\omega)$ is $\tau$-invariant: $g(\tau^{s} \omega)= g(\omega)$. If in addtion, the 
functions $g_{n}$ are ergodic, it follows that the limit $g(\omega)$ is a.s. non random.
\end{proposition}

\section{Generalized Bogoliubov c-numbers approximation}\label{Section-convergence-bogoliubov-pressure}
\setcounter{equation}{0}
\renewcommand{\theequation}{\arabic{section}.\arabic{equation}}
\subsection{Existence of the approximating pressure}\label{appr-press}
Following the Bogoliubov approximation philosophy, we want to replace all creation/annihilation operators
in the momentum states $\psi_{k}$ with kinetic energy less than some $\delta > 0$ by $c$-numbers.
To this end let $I_{\delta} \subset \Lambda_{l}^{*}$ be the set of all \emph{replaceable} modes, that is,
\begin{eqnarray*}
I_{\delta} \, := \, \big\{k \in \Lambda_{l}^{*}: \, k^{2}/2 \leqslant \delta \big\},
\end{eqnarray*}
and we denote $n_{\delta} := \sharp \{k: k \in I_{\delta}\}$. The number of quantum states $n_{\delta}$ is of the order
$V_{l}$, since by definition of the IDS (\ref{density-states-kinetic-energy-operator}):
$n_{\delta} = V_{l} \  \nu_{l}^{0}(\delta)$. Let  $\mathscr{H}_{l}^{\delta}$ to be the subspace of $\mathscr{H}_{l}$
spanned by the set of $\psi_{k}^{l}$ with $k \in I_{\delta}$, and $P_{\delta}$ be orthogonal projector onto this 
subspace.
Hence, we have a natural decomposition of the total space $\mathscr{H}_{l}$ and the corresponding representation 
for the
associated symmetrised Fock space:
\begin{eqnarray}\label{Fock}
\mathscr{H}_{l} = \mathscr{H}_{l}^{\delta} \oplus  \mathscr{H}_{l}^{\perp}\ , \qquad \mathscr{F}_{l} \approx
\mathscr{F}_{l}^{\delta} \otimes  \mathscr{F}_{l}^{\perp}.
\end{eqnarray}
Then we proceed to the Bogoliubov substitution $a_{k} \rightarrow c_{k}$ and $a_{k}^{*} \rightarrow \overline{c}_{k}$
for all $k \in I_{\delta}$, which provides an approximating Hamiltonian that we denote by $H_{l}^{low}(\mu, \{c_{k}\})$.
The meaning of the superscript $low$ will become clear in the next section. We postpone to Appendix A a description of
the explicit form of this operator. The partition function and the corresponding pressure for this approximating
Hamiltonian have the form:
\begin{eqnarray}\label{def-lower-partition-function}
\Xi_{l}^{low} (\mu, \{c_{k}\})  &:=& {\rm{Tr}}_{ \mathscr{F}_{l}^{\perp}} \,\,
e^{-\beta H_{l}^{low}(\mu, \{c_{k}\})} \ ,\\
p_{l,\delta}^{low} (\mu, \{c_{k}\}) \, &:=& \, \frac{1}{V_{l}} \ln \Xi_{l}^{low}(\mu, \{c_{k}\}) \ .
\label{def-lower-pressure}
\end{eqnarray}

The principal result of the present paper is the following main theorem:

\begin{theorem}\label{convergence-approximated-pressure}
The c-numbers substitution for all operators in the energy-band $I_{\delta}$ does not affect the original pressure
(\ref{press}) in the following sense:
\begin{eqnarray}\label{two-limits}
\mathrm{a.s.-}\lim_{l \rightarrow \infty} p_{l}(\beta,\mu) \, = \,
\lim_{\delta \downarrow 0} \liminf_{l \rightarrow \infty} \max_{\{c_{k}\}} \, p_{l,\delta}^{low} (\mu, \{c_{k}\}) 
\, = \, \lim_{\delta \downarrow 0} \limsup_{l \rightarrow \infty} \max_{\{c_{k}\}} \, p_{l,\delta}^{low} 
(\mu, \{c_{k}\}) \ .
\end{eqnarray}
Note that the number of the substituted modes is of order $V$, since we let $\delta \downarrow 0$ after the
thermodynamic limit.
\end{theorem}

\subsection{Proof of the main Theorem}
Our method is a generalisation of the one invented in \cite{LSY}. For the convenience of the reader,
we postpone the proof of some technical lemmas to the next section.

First we define a family of \textit{normalized} coherent vectors in the Fock space $\mathscr{F}_{l}$, with the vacuum
state $\vert 0\rangle$:
\begin{eqnarray}\label{coherent-vector-definition}
\vert c\rangle \, = \, \bigotimes_{k\in I_{\delta}} e^{-\vert c_{k}\vert^{2}/2 +c_{k} a^{*}_{k}} \ \vert 0\rangle \ ,
\end{eqnarray}
labeled by the set of complex numbers $\{c_{k}\}_{k\in I_{\delta}}$. With the help of vectors
(\ref{coherent-vector-definition}), we define the \textit{lower symbol} $A^{low}$ for any operator $A$ in
$\mathscr{F}_{l}$ by the partial inner product:
\begin{eqnarray*}
A^{low}(\{c_{k} \}) \, := \, \langle c \vert A \vert c \rangle \ ,
\end{eqnarray*}
which is an operator in $\mathscr{F}_{l}^{\perp}$, see (\ref{Fock}). Similarly, using the tensor structure
of (\ref{Fock}), we can define on $\mathscr{F}_{l}^{\perp}$ the \textit{upper symbol} $A^{up}$ of the operator $A$ in
$\mathscr{F}_{l}$ through the \textit{integral representation}
\begin{eqnarray*}
A \, =: \,  \int_{\mathbb{C}^{n_{\delta}}} \ud^{2}c_{1} \dots \ud^{2}c_{n_{\delta}} \,
A^{up}(\{c_{k} \}) \, \vert c \rangle \langle c \vert \ .
\end{eqnarray*}
Here $d\, ^{2} c_{j}:= d\,{\rm{Re}}(c_{j})d\,{\rm{Im}}(c_{j})/\pi$ and $\vert c \rangle \langle c \vert :=
\bigotimes_{k\in I_{\delta}}\vert c_{k} \rangle \langle c_{k} \vert$ is the projector on the coherent vectors
(\ref{coherent-vector-definition}). Similar to the one-mode case one has the completeness property
$\int_{\mathbb{C}^{n_{\delta}}} \ud^{2}c_{1} \dots \ud^{2}c_{n_{\delta}} \, \vert c \rangle \langle c \vert = I$.
Note that, contrary to the lower symbols, the upper symbol does not
necessarily \textit{exists}, and it may not be \textit{unique}. Although, this poses no problem in our case,
since the existence (though not the unicity) of the upper symbols follows from the fact that
the Hamiltonian (\ref{multi-part-interacting-Hamiltonian}) is polynomial in
creation/annihilation operators.
We postpone the explicit expressions of these upper symbols to Appendix
\ref{Appendix-explicit-expression-lower-upper-symbol}.

We then define two approximating Hamiltonians, that we denote by $H_{l}^{\textrm{low}} (\mu, \{c_{k}\})$ and
$H_{l}^{\textrm{up}}(\mu, \{c_{k}\})$. They are obtained by replacing all creations and annihilations operators
$a^{\sharp}_{k}$ by their \textit{lower}, respectively \textit{upper}, symbols. We refer the reader to
Appendix \ref{Appendix-explicit-expression-lower-upper-symbol} for the details and the explicit expressions.\\
Note that $H_{l}^{\textrm{low}} (\mu, \{c_{k}\})$ is obtained simply by replacing all operators 
$\{a_{k}^{\sharp}\}_{k\in I_{\delta}}$ by the corresponding complex numbers $\{c_{k}^{\sharp}\}_{k \in I_{\delta}}$. 
Formally it corresponds to the Hamiltonian obtained by the standard Bogoliubov approximation, that is why it 
appears in Theorem \ref{convergence-approximated-pressure}.

In a way similar to (\ref{def-lower-partition-function}), (\ref{def-lower-pressure}),
one can define by $\Xi_{l}^{up}(\mu, \{c_{k}\})$ the partition function for the Hamiltonian
$H_{l}^{\textrm{up}} (\mu, \{c_{k}\})$, and by $p_{l,\delta}^{up}(\mu, \{c_{k}\})$ the corresponding
pressure.

Finally, we denote by $\langle - \rangle_{low}$ and $\langle - \rangle_{up}$ the grand-canonical equilibrium states
related to the following (integrated) partition functions:
\begin{eqnarray}\label{integ-partf-low}
\Xi_{l}^{low}(\mu) \, := \,  \int_{\mathbb{C}^{n_{\delta}}} \ud^{2}c_{1} \dots \ud^{2}c_{n_{\delta}}
{\rm{Tr}}_{ \mathscr{F}_{l}^{\perp}} e^{-\beta H_{l}^{\textrm{low}} (\mu, \{c_{k}\})},\\
\Xi_{l}^{up}(\mu) \, := \,  \int_{\mathbb{C}^{n_{\delta}}} \ud^{2}c_{1} \dots \ud^{2}c_{n_{\delta}}
{\rm{Tr}}_{ \mathscr{F}_{l}^{\perp}} e^{-\beta H_{l}^{\textrm{up}} (\mu, \{c_{k}\})} \ , \label{integ-partf-up}
\end{eqnarray}
and we denote the associated pressures by $p_{l,\delta}^{low}(\mu), p_{l,\delta}^{up}(\mu)$.

Now one can mimic the arguments of \cite{LSY} and extend them to the  \textit{multi-mode} projections
$\vert c \rangle \langle c \vert$ case to produce the Bogoliubov-Peierls and the Berezin-Lieb inequalities for
(\ref{integ-partf-low}) and (\ref{integ-partf-up}). These inequalities 
yield respectively lower and upper estimates for the grand partition function:
\begin{eqnarray}\label{B-L-inequalities}
 \int_{\mathbb{C}^{n_{\delta}}} \ud^{2}c_{1} \dots \ud^{2}c_{n_{\delta}} \,
{\rm{Tr}}_{\mathscr{F}_{l}^{\perp}} e^{-\beta H_{l}^{\textrm{low}} (\mu, \{c_{k}\})} \,
\leqslant \Xi_{l}(\mu)
\leqslant \,  \int_{\mathbb{C}^{n_{\delta}}} \ud^{2}c_{1} \dots \ud^{2}c_{n_{\delta}} \,
{\rm{Tr}}_{\mathscr{F}_{l}^{\perp}} e^{-\beta H_{l}^{\textrm{up}} (\mu, \{c_{k}\})}.
\end{eqnarray}
By a straightforward generalization of the arguments in \cite{LSY} to the case of multi-mode coherent projection
$\vert c \rangle \langle c \vert$, we obtain for any $\{\widetilde{c}_{k}\}_{k\in I_{\delta}}$ the bound
\begin{eqnarray*}
{\rm{Tr}}_{ \mathscr{F}_{l}^{\perp}} e^{-\beta H_{l}^{\textrm{low}} (\mu, \{\widetilde{c}_{k}\})} \, \leqslant \,  
\Xi_{l}(\mu)
\end{eqnarray*}
on the \textit{integrand} in the left-hand side of (\ref{B-L-inequalities}). This in particular implies
\begin{eqnarray}\label{max-pressure-low-symbol-inequality}
\max_{ \{c_{k}\} \in \mathbb{C}^{n_{\delta}}} {\rm{Tr}}_{ \mathscr{F}_{l}^{\perp}} e^{-\beta H_{l}^{\textrm{low}}
(\mu, \{c_{k}\})} \, \leqslant \,  \Xi_{l}(\mu) \ ,
\end{eqnarray}
i.e. the estimate of the grand partition function from \textit{below}.

To find a similar bound for the right-hand side of (\ref{B-L-inequalities}) from \textit{above}, we note that
$H_{l}^{\textrm{low}} (\mu, \{c_{k}\})$  and $H_{l}^{\textrm{up}} (\mu, \{c_{k}\})$ are related by
\begin{eqnarray*}
H_{l}^{\textrm{up}} (\mu, \{c_{k}\}) \, = \, H_{l}^{\textrm{low}} (\mu, \{c_{k}\}) + \kappa(\mu, \{c_{k}\}) \ ,
\end{eqnarray*}
see Appendix \ref{Appendix-explicit-expression-lower-upper-symbol}, equation 
(\ref{appendix-Bogoliubov-expression-kappa}) for an explicit expression of $\kappa(\mu, \{c_{k}\})$. If this 
is combined with the Bogoliubov convexity inequality:
\begin{eqnarray}\label{Bogoliubov-convexity-inequality}
&&\ln  \int_{\mathbb{C}^{n_{\delta}}} \ud^{2}c_{1} \dots \ud^{2}c_{n_{\delta}} {\rm{Tr}}_{ \mathscr{F}_{l}^{\perp}}
e^{-\beta H_{l}^{\textrm{up}} (\mu, \{c_{k}\})} \, - \, \ln  \int_{\mathbb{C}^{n_{\delta}}}
\ud^{2}c_{1} \dots \ud^{2}c_{n_{\delta}} {\rm{Tr}}_{ \mathscr{F}_{l}^{\perp}} e^{-\beta H_{l}^{\textrm{low}}
(\mu, \{c_{k}\})}
\nonumber \\
&\leqslant& \frac{  \int_{\mathbb{C}^{n_{\delta}}} \ud^{2}c_{1} \dots \ud^{2}c_{n_{\delta}}
{\rm{Tr}}_{ \mathscr{F}_{l}^{\perp}} \big(-\kappa(\mu,
\{c_{k}\}) e^{-\beta H_{l}^{\textrm{up}} (\mu, \{c_{k}\})}\big)}{ \int_{\mathbb{C}^{n_{\delta}}} \ud^{2}c_{1}
\dots \ud^{2}c_{n_{\delta}}
{\rm{Tr}}_{ \mathscr{F}_{l}^{\perp}} e^{-\beta H_{l}^{\textrm{up}} (\mu, \{c_{k}\})}}
\label{Bogoliubov-convexity-inequality}\ ,
\end{eqnarray}
then (\ref{B-L-inequalities}) and (\ref{Bogoliubov-convexity-inequality}) provide the upper bound:
\begin{eqnarray}\label{upper-bound-1-true-partition-function}
\ln {\rm{Tr}}_{ \mathscr{F}_{l}} e^{-\beta H_{l}(\mu)} \, \leqslant \, \ln  \int_{\mathbb{C}^{n_{\delta}}}
\ud^{2}c_{1} \dots \ud^{2}c_{n_{\delta}} {\rm{Tr}}_{ \mathscr{F}_{l}^{\perp}} e^{-\beta H_{l}^{\textrm{low}}
(\mu, \{c_{k}\})} \, - \, \langle \kappa(\mu, \{c_{k}\}) \rangle_{up} \ .
\end{eqnarray}
Using the orthogonal projection $P_{\delta}: \mathscr{H}_{l} \mapsto \mathscr{H}_{l}^{\delta}$, and in view of
(\ref{appendix-Bogoliubov-expression-kappa}), one can estimate the last term in
(\ref{upper-bound-1-true-partition-function}) explicitly:
{\allowdisplaybreaks
\begin{eqnarray}\label{upper-bound-difference-up-low-hamiltonians}
-  \kappa(\mu, \{c_{k}\}) &\leqslant& {\rm{Tr}} (h_{l}^{\omega} - \mu) P_{\delta}\\
&-& \gamma \Big( \nu_{l}^{0}(\delta) + \frac{V_{l}}{2} \big(\nu_{l}^{0}(\delta)\big)\, ^{2} + V_{l} 
\nu_{l}^{0}(\delta)
\nu_{l}^{0}(2\delta)\Big) \nonumber\\
&+& \frac{\gamma}{2} \big( \frac{4}{V_{l}} + 2 \nu_{l}^{0}(\delta) + 2  \nu_{l}^{0}(2\delta) \big)
\sum_{k \in I_{\delta}}
\vert c_{k}\vert\, ^{2}\nonumber\\
&+&\frac{\gamma}{2} \big( 2 \nu_{l}^{0}(\delta) + 2  \nu_{l}^{0}(\delta) \big)\sum_{k \in I_{\delta}^{c}}
a^{*}_{k} a_{k}\nonumber\\
&\leqslant&  {\rm{Tr}} \big((h_{l}^{\omega} - \mu) P_{\delta}\big) - \gamma  \nu_{l}^{0}(\delta) 
\Big( 1 - 4 V_{l} \nu_{l}^{0}(2\delta) +
\frac{V_{l}}{2} \nu_{l}^{0}(\delta) +  V_{l} \nu_{l}^{0}(2\delta)\Big) \nonumber\\
&+& 4 \gamma \nu_{l}^{0}(2\delta) \Big( \sum_{k \in I_{\delta}} \big(\vert c_{k}\vert\, ^{2} - 1\big) +
\sum_{k \in I_{\delta}^{c}} a^{*}_{k} a_{k} \Big)\nonumber \ .
\end{eqnarray}}
Here $I_{\delta}^{c}:= \Lambda_{l}^{*}\setminus I_{\delta}$. Then taking into account the upper symbol of the total
number operator, we find that
\begin{eqnarray}\label{Upper-symbol-particle-number-operator}
H_{l}^{\textrm{up}} (\mu, \{c_{k}\})  + a\Big( \sum_{k \in I_{\delta}} (\vert c_{k}\vert\, ^{2} - 1) +
\sum_{k \in I_{\delta}^{c}} a^{*}_{k} a_{k}\Big)
\, = \, H_{l}^{\textrm{up}} (\mu - a, \{c_{k}\}) \ ,
\end{eqnarray}
which together with equation (\ref{upper-bound-1-true-partition-function}) provides the following estimate
\begin{eqnarray}\label{upper-bound-1-true-pressure}
\ln {\rm{Tr}}_{ \mathscr{F}_{l}} e^{-\beta H_{l}(\mu)}  &\leqslant& \ln  \int_{\mathbb{C}^{n_{\delta}}}
\ud^{2}c_{1} \dots \ud^{2}c_{n_{\delta}} {\rm{Tr}}_{ \mathscr{F}_{l}^{\perp}} e^{-\beta H_{l}^{\textrm{low}}
(\mu, \{c_{k}\})} \\
&+&{\rm{Tr}} \big((h_{l}^{\omega} - \mu) P_{\delta}\big)  - \gamma  \nu_{l}^{0}(\delta) \Big( 1 - 4 V_{l} 
\nu_{l}^{0}(2\delta) + \frac{V_{l}}{2} \nu_{l}^{0}(\delta) +  V_{l} \nu_{l}^{0}(2\delta)\Big)\nonumber\\
&+& 4 \gamma \nu_{l}^{0}(2\delta) \partial_{\mu} \ln \int_{\mathbb{C}^{n_{\delta}}} \ud^{2}c_{1} 
\dots \ud^{2}c_{n_{\delta}} {\rm{Tr}}_{ \mathscr{F}_{l}^{\perp}} e^{-\beta H_{l}^{\textrm{low}} 
(\mu, \{c_{k}\})}\nonumber\\
\end{eqnarray}
To finish the proof, we need three lemmas. We postpone their proofs to the next section.
\begin{lemma}\label{lemma-upper-bound-integrated-particle-density-upper-symbol}
Suppose that the system of interacting bosons (\ref{multi-part-interacting-Hamiltonian}) has a bounded
limiting particle density $\rho(\mu)$ for any fixed $\mu \in \mathbb{R}$:
\begin{equation}\label{part-dens}
\rho(\mu):= \partial_{\mu} p(\mu):= \partial_{\mu} \lim_{l \rightarrow \infty} p_{l}(\beta,\mu) \, < \, \infty
\end{equation}
see (\ref{press}). Then one gets the estimates:
\begin{eqnarray}\label{upper-bound-integrated-particle-density-lower-symbol}
\limsup_{\delta \downarrow 0} \limsup_{l \rightarrow \infty} \frac{1}{\beta V_{l}}
\partial_{\mu} p_{l,\delta}^{low}(\mu) \,
\, \leqslant \, \, \rho(\mu),\\
\limsup_{\delta \downarrow 0} \limsup_{l \rightarrow \infty} \frac{1}{\beta V_{l}}
\partial_{\mu} p_{l,\delta}^{up}(\mu) \,
\, \leqslant \, \,  \rho(\mu) \ . \label{upper-bound-integrated-particle-density-upper-symbol}
\end{eqnarray}
\end{lemma}

Next, we relate the integrated pressure $p_{l,\delta}^{low}(\mu)$ defined by (\ref{integ-partf-low})
to the maximum of the corresponding integrand.
\begin{lemma}\label{lemma-upper-bound-integrated-pressure-in-term-max-pressure}
For any $\alpha > 1$, one has the estimate:
\begin{eqnarray}\label{upper-bound-integrated-pressure-in-term-max-pressure}
&&\frac{1}{\beta V_{l}} \ln \int_{\mathbb{C}^{n_{\delta}}} \ud^{2}c_{1} \dots \ud^{2}c_{n_{\delta}}
{\rm{Tr}}_{\mathscr{F}_{l}^{\perp}} e^{-\beta H_{l}^{\textrm{low}} (\mu, \{c_{k}\})}\\
&\leqslant& \frac{1}{\beta V_{l}} \ln \max_{\{c_{k}\}} {\rm{Tr}}_{\mathscr{F}_{l}^{\perp}} 
e^{-\beta H_{l}^{\textrm{low}}(\mu, \{c_{k}\})} \, -
\, \frac{1}{\beta V_{l}} \ln (1-1/\alpha) \, + \, \frac{\nu_{l}^{0}(\delta)}{\beta} \ln (\alpha \partial_{\mu}
p_{l,\delta}^{low}(\mu))\nonumber\\
&+& \frac{1}{\beta} \nu_{l}^{0}(\delta) -  \frac{1}{2\beta} \frac{\ln V_{l}}{V_{l}} - 
\frac{\nu_{l}^{0}(\delta)}{\beta} \ln \big(\nu_{l}^{0}(\delta)\big) - \frac{1}{2\beta V_{l}} 
\ln \big(\nu_{l}^{0}(\delta)\big). \nonumber
\end{eqnarray}
\end{lemma}

Notice that above statements are independent of the possible presence of random potentials.
The next Lemma serves to control (ergodic) random external potentials.
\begin{lemma}\label{lemma-upper-bound-random-potential-correction}
Taking into account our assumptions on the random potentials, see Section \ref{Section-models-definitions},
one gets the following inequality:
\begin{eqnarray*}
\limsup_{l \rightarrow \infty}\frac{1}{V_{l}} {\rm{Tr}} (h_{l}^{\omega} - \mu)P_{\delta} \, \leqslant \,
\nu^{0}(\delta) \,
\Big( (\delta - \mu) + \mathbb{E}\big( v^{\omega}(0) \big)  \Big) \ .
\end{eqnarray*}
Here $\mathbb{E}(\cdot)$ denotes the expectations in the probability space $(\Omega, \mathcal{F}, \mathbb{P})$.
\end{lemma}

Returning back to the proof of Theorem \ref{convergence-approximated-pressure}, we get from
(\ref{max-pressure-low-symbol-inequality}) and (\ref{upper-bound-1-true-pressure}) the estimates:
\begin{eqnarray*}
&&\max_{\{c_{k}\}} \, p_{l,\delta}^{low} (\mu, \{c_{k}\}) \leqslant p_{l}(\mu) \leqslant p_{l,\delta}^{low} (\mu) +
\frac{1}{\beta V_{l}} \Big({\rm{Tr}} \big((h_{l}^{\omega} - \mu) P_{\delta}\big)  -
\gamma  \nu_{l}^{0}(\delta) \big( 1 - 4 V_{l} \nu_{l}^{0}(2\delta) + \\
&&\frac{V_{l}}{2} \nu_{l}^{0}(\delta) +
V_{l} \nu_{l}^{0}(2\delta)\big)\Big) + 4 \gamma \nu_{l}^{0}(2\delta) \frac{1}{\beta V_{l}} \partial_{\mu}
\ln \int_{\mathbb{C}^{n_{\delta}}} \ud^{2}c_{1} \dots \ud^{2}c_{n_{\delta}} {\rm{Tr}}_{ \mathscr{F}_{l}^{\perp}}
e^{-\beta H_{l}^{\textrm{low}}(\mu, \{c_{k}\})} \ . \nonumber
\end{eqnarray*}
By Lemma \ref{lemma-upper-bound-integrated-pressure-in-term-max-pressure}, this implies for any configuration
$\omega$ the estimates:
\begin{eqnarray}\label{estimate-max.p-less-true.p-less_max.p+correction}
\max_{\{c_{k}\}} \, p_{l,\delta}^{low} (\mu, \{c_{k}\}) \, \leqslant \, p_{l}(\mu) \, \leqslant \, \max_{\{c_{k}\}} \,
p_{l,\delta}^{low} (\mu, \{c_{k}\}) \, + \, K^{{\omega}}(l,\delta) \ ,
\end{eqnarray}
where the random parameter $K^{\omega}(l,\delta)$ is given by
\begin{eqnarray*}
 K^{\omega}(l,\delta) :&=& \frac{1}{\beta V_{l}} \Big({\rm{Tr}} \big((h_{l}^{\omega} - \mu) P_{\delta}\big)  - \gamma
 \nu_{l}^{0}(\delta) \big( 1 - 4 V_{l} \nu_{l}^{0}(2\delta) + \frac{V_{l}}{2} \nu_{l}^{0}(\delta) +  
 V_{l} \nu_{l}^{0}(2\delta)\big)\Big)\\
&+& 4 \gamma \nu_{l}^{0}(2\delta) \frac{1}{\beta V_{l}} \partial_{\mu} \ln \int_{\mathbb{C}^{n_{\delta}}}
\ud^{2}c_{1} \dots \ud^{2}c_{n_{\delta}} {\rm{Tr}}_{ \mathscr{F}_{l}^{\perp}} e^{-\beta H_{l}^{\textrm{low}}
(\mu, \{c_{k}\})}.
\end{eqnarray*}
Note that, by Lemmas \ref{lemma-upper-bound-integrated-particle-density-upper-symbol} and
\ref{lemma-upper-bound-random-potential-correction}, we can control this error term since for any configuration
one gets:
\begin{eqnarray*}
\lim_{\delta \downarrow 0} \liminf_{l \rightarrow \infty} K^{\omega}(l,\delta) \, = \,
\lim_{\delta \downarrow 0} \limsup_{l \rightarrow \infty} K^{\omega}(l,\delta) \, = \, 0 \ .
\end{eqnarray*}
Hence, (\ref{estimate-max.p-less-true.p-less_max.p+correction}) yields
\begin{eqnarray*}
\limsup_{l \rightarrow \infty} \, \max_{\{c_{k}\}} \, p_{l,\delta}^{low} (\mu, \{c_{k}\}) \,
\leqslant \, p(\mu) \, \leqslant \,
\limsup_{l \rightarrow \infty} \, \max_{\{c_{k}\}} \, p_{l,\delta}^{low} (\mu, \{c_{k}\}) \, +
\liminf_{l \rightarrow \infty} \,
\, K^{\omega}(l,\delta)\\
\limsup_{\delta \downarrow 0} \, \limsup_{l \rightarrow \infty} \, \max_{\{c_{k}\}} \, p_{l,\delta}^{low}
(\mu, \{c_{k}\}) \,
\leqslant \, p(\mu) \, \leqslant \, \, \liminf_{\delta \downarrow 0} \limsup_{l \rightarrow \infty} \,
\max_{\{c_{k}\}} \,
p_{l,\delta}^{low} (\mu, \{c_{k}\}) \ ,
\end{eqnarray*}
which proves the first equality in (\ref{two-limits}) in Theorem \ref{convergence-approximated-pressure}.
The second one can be proven in a similar way.

\subsection{Proofs of technical results}

Recall that the \textit{finite-volume} IDS (\ref{density-states-kinetic-energy-operator}): $\nu_{l}^{0}(\delta)\geq 0$,
converges in the limit $l \rightarrow \infty$ to the Weyl formula $\nu^{0}(\delta) = C_{d} \delta^{d/2}$, 
uniformly in $\delta$
on any finite interval  $0 \leq \delta \leq E$. Below we use this uniformity in a systematic way.

\noindent \textbf{Proof of Lemma \ref{lemma-upper-bound-integrated-particle-density-upper-symbol}}\\
Since
\begin{eqnarray}\label{Lower-symbol-particle-number-operator}
H_{l}^{\textrm{low}} (\mu, \{c_{k}\})  + a\Big( \sum_{k \in I_{\delta}} \vert c_{k}\vert\, ^{2} + 
\sum_{k \in I_{\delta}^{c}} a^{*}_{k} a_{k}\Big) \, = \, H_{l}^{\textrm{low}} (\mu - a, \{c_{k}\}) \ ,
\end{eqnarray}
the estimate (\ref{upper-bound-difference-up-low-hamiltonians}) yields:
\begin{eqnarray*}
H_{l}^{\textrm{up}} (\mu, \{c_{k}\}) &\geqslant& H_{l}^{\textrm{low}} (\mu + 4\gamma \nu_{l}^{0}(2\delta), \{c_{k}\})\\
&-&  {\rm{Tr}} \big((h_{l}^{\omega} - \mu) P_{\delta}\big) - \gamma  \nu_{l}^{0}(\delta) \Big( 1 + \frac{V_{l}}{2}
\nu_{l}^{0}(\delta) +  \nu_{l}^{0}(2\delta)\Big).
\end{eqnarray*}
Applying now the Bogoliubov-Peierls and the Berezin-Lieb inequalities, we obtain:
\begin{eqnarray*}
p_{l,\delta}^{low}(\mu) \, \leqslant \, p_{l}(\mu) \, \leqslant \, p_{l,\delta}^{up}(\mu) \, \leqslant \,
p_{l,\delta}^{low}
(\mu + 4\gamma \nu_{l}^{0}(2\delta)) + \frac{M^{\omega}(l,\delta)}{V_{l}} \ ,
\end{eqnarray*}
where
\begin{eqnarray} \label{M1}
M^{\omega}(l,\delta) \, := \,  {\rm{Tr}} \big((h_{l}^{\omega} - \mu) P_{\delta}\big) + \gamma  \nu_{l}^{0}(\delta)
\Big( 1 +
\frac{V_{l}}{2} \nu_{l}^{0}(\delta) +  \nu_{l}^{0}(2\delta)\Big) \ .
\end{eqnarray}
Consequently, for any configuration $\omega$ one gets in the limit the estimates:
\begin{eqnarray}
\limsup_{l \rightarrow \infty} p_{l,\delta}^{low}(\mu) \leqslant p(\mu) \leqslant \liminf_{l \rightarrow \infty}
p_{l,\delta}^{up}(\mu) \leqslant \limsup_{l \rightarrow \infty} p_{l,\delta}^{up}(\mu)
&\leqslant&  \liminf_{l \rightarrow \infty} p_{l,\delta}^{low}(\mu + 4\gamma \nu_{l}^{0}(2\delta)) \nonumber \\
&+&\liminf_{l \rightarrow \infty} \frac{M^{\omega}(l,\delta)}{V_{l}} \ . \label{estim 1}
\end{eqnarray}
Since $p_{l,\delta}^{up}(\mu)$ is a convex functions of $\mu$, for any $t > 0$ we have
\begin{eqnarray*}
\partial_{\mu} p_{l,\delta}^{up}(\mu) \, \leqslant \, \frac{1}{t}\big( p_{l,\delta}^{up}(\mu + t) -
p_{l,\delta}^{up}(\mu)\big) \ .
\end{eqnarray*}
Then by virtue of estimates (\ref{estim 1}) we obtain:
\begin{eqnarray}\label{M2}
\limsup_{l \rightarrow \infty} \partial_{\mu} p_{l,\delta}^{up}(\mu)  &\leqslant& \frac{1}{t}
\big(\limsup_{l \rightarrow \infty}
p_{l,\delta}^{up}(\mu + t) - \liminf_{l \rightarrow \infty}p_{l,\delta}^{up}(\mu)\big)\\ \nonumber
&\leqslant&\frac{1}{t}\big(\liminf_{l \rightarrow \infty} p_{l,\delta}^{low}(\mu + t +
4\gamma \nu_{l}^{0}(2\delta)) + \liminf_{l \rightarrow \infty} \frac{1}{V_{l}} M^{\omega}(l,\delta,\mu) -
p(\mu)\big)\\ \nonumber
&\leqslant&\frac{1}{t}\big( p(\mu + t + 4\gamma \nu^{0}(2\delta))
+ \liminf_{l \rightarrow \infty} \frac{1}{V_{l}} M^{\omega}(l,\delta,\mu) - p(\mu)\big) \ . \nonumber
\end{eqnarray}
Since by (\ref{M1}) and Lemma \ref{lemma-upper-bound-random-potential-correction}
for any configuration $\omega$ one gets:
\begin{eqnarray*}
\lim_{\delta \downarrow 0} \, \liminf_{l \rightarrow \infty} \frac{1}{V_{l}} M^{\omega}(l,\delta) \, = \, 0 \ ,
\end{eqnarray*}
we obtain from (\ref{M2})
\begin{eqnarray}\label{proof-lemma-Bogoliubov-estimate-density-upper-symbol}
\limsup_{\delta \downarrow 0}\limsup_{l \rightarrow \infty} \partial_{\mu} p_{l,\delta}^{up}(\mu)
\, \leqslant \, \frac{1}{t}
\big( p(\mu + t) - p(\mu)\big)
\end{eqnarray}
which is valid for any $t > 0$. Letting $t \downarrow 0$ leads to the second estimate
(\ref{upper-bound-integrated-particle-density-upper-symbol}).

Similarly one proves the first estimate (\ref{upper-bound-integrated-particle-density-lower-symbol}).
Again using the convexity and (\ref{estim 1}) we obtain
\begin{eqnarray*}
\limsup_{l \rightarrow \infty} \partial_{\mu} p_{l,\delta}^{low}(\mu) &\leqslant&
\frac{1}{t}\big(\limsup_{l \rightarrow \infty} p_{l,\delta}^{low}(\mu + t) -
\liminf_{l \rightarrow \infty} p_{l,\delta}^{low}(\mu)\big)\\
&\leqslant& \frac{1}{t}\big(\limsup_{l \rightarrow \infty} p_{l}(\mu + t)\\
&-& \limsup_{l \rightarrow \infty}
p_{l,\delta}^{up}(\mu - 4\gamma \nu_{l}^{0}(2\delta))
- \liminf_{l \rightarrow \infty} \frac{1}{V_{l}} M^{\omega}(l,\delta,\mu - 4\gamma \nu_{l}^{0}(2\delta)) \big) \ .
\end{eqnarray*}
Applying the estimate (\ref{estim 1}) to the last inequality once more, we get
\begin{eqnarray*}
\limsup_{l \rightarrow \infty} \partial_{\mu} p_{l,\delta}^{low}(\mu)
&\leqslant& \frac{1}{t}\big(p(\mu + t)- p(\mu - 4\gamma \nu^{0}(2\delta))
- \liminf_{l \rightarrow \infty} \frac{1}{V_{l}} M^{\omega}(l,\delta,\mu - 4\gamma \nu_{l}^{0}(2\delta)) \big)
\end{eqnarray*}
and in view of the limit (\ref{M1}) and Lemma \ref{lemma-upper-bound-random-potential-correction}, 
the result follows by
letting $\delta \downarrow 0$ and then, letting $t \downarrow 0$.  \hfill $\square$

\noindent \textbf{Proof of Lemma \ref{lemma-upper-bound-integrated-pressure-in-term-max-pressure}}\\
Let
\begin{eqnarray*}
\mathbb{C}^{n_{\delta}}_{\xi} \, := \, \{z \in \mathbb{C}^{n_{\delta}}: \vert z \vert\, ^{2} \leqslant \xi\}.
\end{eqnarray*}
and we denote the volume of this ball by $\textrm{Vol}(\mathbb{C}^{n_{\delta}}_{\xi})= 
\pi^{n_{\delta}}\xi^{n_{\delta}}/{n_{\delta}\Gamma(n_{\delta})}$.

We then obtain the following bound
\begin{eqnarray*}
&&\int_{\mathbb{C}^{n_{\delta}}} \ud^{2}c_{1} \dots \ud^{2}c_{n_{\delta}} {\rm{Tr}}_{ \mathscr{F}_{l}^{\perp}}
e^{-\beta H_{l}^{\textrm{low}} (\mu, \{c_{k}\})}\\
&=& \int_{\mathbb{C}^{n_{\delta}}_{\xi}} \ud^{2}c_{1} \dots \ud^{2}c_{n_{\delta}} {\rm{Tr}}_{ \mathscr{F}_{l}^{\perp}}
e^{-\beta H_{l}^{\textrm{low}} (\mu, \{c_{k}\})} \, + \, \int_{\mathbb{C}^{n_{\delta}}
\symbol{92}
\mathbb{C}^{n_{\delta}}_{\xi} } \ud^{2}c_{1} \dots \ud^{2}c_{n_{\delta}} {\rm{Tr}}_{ \mathscr{F}_{l}^{\perp}}
e^{-\beta H_{l}^{\textrm{low}}
(\mu, \{c_{k}\})}\\
&\leqslant& \frac{\textrm{Vol}(\mathbb{C}^{n_{\delta}}_{\xi})}{\pi^{n_{\delta}}} \max_{\{c_{k}\}} 
{\rm{Tr}}_{ \mathscr{F}_{l}^{\perp}}
e^{-\beta H_{l}^{\textrm{low}}
(\mu, \{c_{k}\})} \, + \, \frac{1}{\xi\pi^{n_{\delta}}}  \int_{\mathbb{C}^{n_{\delta}}} \ud^{2}c_{1} \dots \ud^{2}
c_{n_{\delta}}
\big(\sum_{k \in I_{\delta}}\vert c_{k} \vert\, ^{2}\big) {\rm{Tr}}_{ \mathscr{F}_{l}^{\perp}}
e^{-\beta H_{l}^{\textrm{low}} (\mu, \{c_{k}\})}\\
&\leqslant& \frac{\textrm{Vol}(\mathbb{C}^{n_{\delta}}_{\xi})}{\pi^{n_{\delta}}} \max_{\{c_{k}\}} 
{\rm{Tr}}_{ \mathscr{F}_{l}^{\perp}}
e^{-\beta H_{l}^{\textrm{low}}
(\mu, \{c_{k}\})} \, + \, \frac{1}{\xi\pi^{n_{\delta}}} \,\,\langle \sum_{k \in I_{\delta}}\vert c_{k} \vert\, ^{2}
\rangle_{low} \,\,\int_{\mathbb{C}^{n_{\delta}}} \ud^{2}c_{1} \dots \ud^{2}c_{n_{\delta}}
{\rm{Tr}}_{ \mathscr{F}_{l}^{\perp}} e^{-\beta H_{l}^{\textrm{low}} (\mu, \{c_{k}\})}.
\end{eqnarray*}
Here we used the expectation $\langle \ \cdot \ \rangle_{low}$ defined by the integrated partition function
$\Xi_{l}^{low}(\mu)$, see (\ref{integ-partf-low}). Notice that, using the explicit form of the lower symbol
for the total particle number operator,
see (\ref{Lower-symbol-particle-number-operator}), we can represent the last integral in the form:
\begin{eqnarray*}
&&\int_{\mathbb{C}^{n_{\delta}}} \ud^{2}c_{1} \dots \ud^{2}c_{n_{\delta}} {\rm{Tr}}_{ \mathscr{F}_{l}^{\perp}}
e^{-\beta H_{l}^{\textrm{low}} (\mu, \{c_{k}\})}\\
&\leqslant& \frac{\textrm{Vol}(\mathbb{C}^{n_{\delta}}_{\xi})}{\pi^{n_{\delta}}} \max_{\{c_{k}\}} 
{\rm{Tr}}_{ \mathscr{F}_{l}^{\perp}}
e^{-\beta H_{l}^{\textrm{low}}
(\mu, \{c_{k}\})} \, + \, \frac{1}{\xi\pi^{n_{\delta}}} \big(V_{l} \partial_{\mu} p_{l,\delta}^{low}(\mu) \big)
\int_{\mathbb{C}^{n_{\delta}}} \ud^{2}c_{1} \dots \ud^{2}c_{n_{\delta}} {\rm{Tr}}_{ \mathscr{F}_{l}^{\perp}}
e^{-\beta H_{l}^{\textrm{low}}
(\mu, \{c_{k}\})}\\
&=& \big(\frac{\xi^{n_{\delta}}}{n_{\delta}\Gamma(n_{\delta})} \big)  \max_{\{c_{k}\}}
{\rm{Tr}}_{ \mathscr{F}_{l}^{\perp}}
e^{-\beta H_{l}^{\textrm{low}} (\mu, \{c_{k}\})} \, + \, \frac{1}{\xi\pi^{n_{\delta}}} \big( V_{l}
\partial_{\mu} p_{l,\delta}^{low}(\mu) \big)\int_{\mathbb{C}^{n_{\delta}}} \ud^{2}c_{1} \dots \ud^{2}c_{n_{\delta}}
{\rm{Tr}}_{ \mathscr{F}_{l}^{\perp}} e^{-\beta H_{l}^{\textrm{low}} (\mu, \{c_{k}\})} \ .
\end{eqnarray*}
Thus, we get that
\begin{eqnarray*}
\big(1-\frac{V_{l}}{\xi}  \partial_{\mu} p_{l,\delta}^{low}(\mu)\big)
\int_{\mathbb{C}^{n_{\delta}}}
\ud^{2}c_{1} \dots \ud^{2}c_{n_{\delta}} {\rm{Tr}}_{ \mathscr{F}_{l}^{\perp}} e^{-\beta H_{l}^{\textrm{low}}
(\mu, \{c_{k}\})} \, \leqslant \,
\big(\frac{\xi^{n_{\delta}}}{n_{\delta}\Gamma(n_{\delta})} \big)  \max_{\{c_{k}\}} 
{\rm{Tr}}_{ \mathscr{F}_{l}^{\perp}}
e^{-\beta H_{l}^{\textrm{low}}
(\mu, \{c_{k}\})} \ .
\end{eqnarray*}
If we put $\xi := \alpha V_{l} \partial_{\mu} p_{l,\delta}^{low}(\mu)$ for some $\alpha > 1$, and use the
Stirling formula for large $n_{\delta}$, then the coefficient in the right-hand side can be estimated as:
\begin{eqnarray*}
\frac{\xi^{n_{\delta}}}{n_{\delta}\Gamma(n_{\delta})} \, \leqslant \, \frac{\big(\alpha V_{l}  \partial_{\mu}
p_{l,\delta}^{low}(\mu)\big)^{n_{\delta}}}{n_{\delta} n_{\delta}^{n_{\delta}-1/2} e^{-n_{\delta}}} \, \leqslant \,
\frac{\Big( \big(\alpha  \partial_{\mu} p_{l,\delta}^{low}(\mu)\big)^{\nu_{l}^{0}(\delta)}\Big)^{V_{l}}}
{\big(V_{l} \nu_{l}^{0}(\delta)\big)^{V_{l} \nu_{l}^{0}(\delta) +1/2}} \, e^{-n_{\delta}} \ .
\end{eqnarray*}
Hence, one finally obtains the estimate:
\begin{eqnarray*}
&&\int_{\mathbb{C}^{n_{\delta}}} \ud^{2}c_{1} \dots \ud^{2}c_{n_{\delta}} {\rm{Tr}}_{ \mathscr{F}_{l}^{\perp}}
e^{-\beta H_{l}^{\textrm{low}} (\mu, \{c_{k}\})}\\
&\leqslant&
\frac{1}{1-{1}/{\alpha}}\Big( \big(\alpha  \partial_{\mu} p_{l,\delta}^{low}(\mu)\big)^{\nu_{l}^{0}
(\delta)}\Big)^{V_{l}} V^{-1/2} \big( \nu_{l}^{0}(\delta) \big)^{-(\nu_{l}^{0}(\delta)V_{l} + 1/2)}
e^{\nu_{l}^{0}(\delta)V_{l}} \, \max_{\{c_{k}\}} {\rm{Tr}}_{ \mathscr{F}_{l}^{\perp}} e^{-\beta H_{l}^{\textrm{low}}
(\mu, \{c_{k}\})},
\end{eqnarray*}
which leads to the desired result
\begin{eqnarray}\label{upper-bound-integrated-pressure-in-term-max-pressure}
&&\frac{1}{\beta V_{l}} \ln \int_{\mathbb{C}^{n_{\delta}}} \ud^{2}c_{1} \dots \ud^{2}c_{n_{\delta}}
{\rm{Tr}}_{ \mathscr{F}_{l}^{\perp}} e^{-\beta H_{l}^{\textrm{low}} (\mu, \{c_{k}\})}\\
&\leqslant& \frac{1}{\beta V_{l}} \ln \max_{\{c_{k}\}} {\rm{Tr}}_{ \mathscr{F}_{l}^{\perp}} 
e^{-\beta H_{l}^{\textrm{low}}
(\mu, \{c_{k}\})}
\, - \,
\frac{1}{\beta V_{l}} \ln (1-1/\alpha) \, + \, \frac{\nu_{l}^{0}(\delta)}{\beta} \ln (\alpha \partial_{\mu}
p_{l,\delta}^{low}(\mu))\nonumber\\
&+& \frac{1}{\beta} \nu_{l}^{0}(\delta) -  \frac{1}{2\beta} \frac{\ln V_{l}}{V_{l}} - 
\frac{\nu_{l}^{0}(\delta)}{\beta}
\ln \big(\nu_{l}^{0}(\delta)\big) - \frac{1}{2\beta V_{l}} \ln \big(\nu_{l}^{0}(\delta)\big). \nonumber
\end{eqnarray}
\hfill $\square$

\noindent \textbf{Proof of Lemma \ref{lemma-upper-bound-random-potential-correction}}\\
First we note that since the projection $P_{\delta}$ is constructed with respect to the basis of eigenvectors of
$h_{l}^{0}$, one gets:
\begin{eqnarray*}
\frac{1}{V_{l}} {\rm{Tr}} (h_{l}^{\omega} - \mu)P_{\delta} &=& \frac{1}{V_{l}} {\rm{Tr}} (h_{l}^{0} - \mu)P_{\delta}
+ \frac{1}{V_{l}} {\rm{Tr}} (v^{\omega}\upharpoonright_{\Lambda_{l}})P_{\delta}\\
&\leqslant& (\delta - \mu) \nu_{l}^{0}(\delta) + \frac{1}{V_{l}} \sum_{k \in I_{\delta}} (\psi_{k}^{l},
v^{\omega} \psi_{k}^{l}) \ ,
\end{eqnarray*}
and consequently:
\begin{eqnarray*}
\frac{1}{V_{l}} {\rm{Tr}} (h_{l}^{\omega} - \mu)P_{\delta} &\leqslant& (\delta - \mu)\nu_{l}^{0}(\delta) \, + \,
\int_{\left[0, \delta\right)} (\psi_{k}^{l}, v^{\omega} \psi_{k}^{l}) \, \nu_{l}^{0}(\ud k)\\
&=& (\delta - \mu)\nu_{l}^{0}(\delta) \, + \,
\int_{\left[0, \delta\right)} \nu_{l}^{0}(\ud k) \, \frac{1}{V_{l}} \int_{\Lambda_{l}} \ud x \ v^{\omega}(x)\\
&=& \nu_{l}^{0}(\delta) \Big( (\delta - \mu)  \, + \, \frac{1}{V_{l}} \int_{\Lambda_{l}} \ud x \ v^{\omega}(x)\Big) \ .
\end{eqnarray*}
Thus, the ergodic theorem yields:
\begin{eqnarray*}
\limsup_{l \rightarrow \infty} \, \frac{1}{V_{l}} {\rm{Tr}} (h_{l}^{\omega} - \mu)P_{\delta} \, \leqslant \,
 \nu^{0}(\delta) \Big( (\delta - \mu)  \, + \, \mathbb{E}_{\omega}\big(v^{\omega}(0)\big)\Big) \ .
\end{eqnarray*}
\hfill $\square$
\section{Concluding remarks} \label{Section-condensation}

We discuss here the meaning of our main result and in particular the way to interpret the solutions
of the variational problem established in Theorem \ref{convergence-approximated-pressure}. Note that the theorem
is established for a random system in the thermodynamic limit. Its aim is to take into account
the possibility of type III generalized condensation, which we believe is favored by the randomness even
in interacting systems.

First we recall a result established in \cite{LSY}, cf. Section 1.1. For a homogeneous system and the single-mode
substitution at $k = 0$, the solution of the variational problem gives the \textit{total} condensate density 
in the mode $k = 0$, \emph{if} one adds to the Hamiltonian the \textit{zero-mode} gauge-breaking term 
(quasi-average sources):
\begin{eqnarray*}
H_{l} (\mu; \eta) \, := \, H_{l} (\mu) \, + \, \sqrt{V_{l}} \ \big( \, \overline{\eta} \, a_{0} +
\eta \, a^{*}_{0} \, \big) \ .
\end{eqnarray*}
This means that after the Bogoliubov $c$-number substitution the solution ${\alpha}_{\max,l}(\beta,\mu; \eta)$
of the (finite-volume) variational problem not only provides the right pressure  in the thermodynamic limit, but 
it also coincides with quasi-averages that give the \textit{total} amount of the condensate in the \textit{zero} 
mode:
\begin{eqnarray*}
\lim_{\vert \eta \vert \downarrow 0} \ \lim_{l \rightarrow \infty} {\alpha}_{\max,l}(\beta,\mu;\eta) \, =
\, \lim_{\vert \eta \vert \downarrow 0} \ \lim_{l \rightarrow \infty} \langle a^{*}_{0}a_{0}/V_l \rangle_{l}
(\beta,\mu; \eta) \ .
\end{eqnarray*}
Here $ \langle -\rangle_{l}(\beta,\mu; \eta)$ is the equilibrium state defined by $H_{l} (\mu; \eta)$.

Using a simple example, we discuss the relevance of this quasi-average approach to more subtle cases of the
condensation of type II and III. We show that the Bogoliubov quasi-average sources breaking the gauge 
invariance \cite{Bog} are able to alter the fine structure of the condensate reducing it to one-mode (type I).

To see this, consider the perfect Bose-gas in a cubic three-dimensional anisotropic parallelepiped
$\Lambda_{l} := V_{l}^{\alpha_x}\times V_{l}^{\alpha_y}\times V_{l}^{\alpha_z}$, with periodic boundary
condition and $\alpha_x \geq \alpha_y \geq \alpha_z$, $\alpha_x + \alpha_y + \alpha_z = 1$. The Hamiltonian
is:
\begin{eqnarray}\label{freeGBEC}
H^{0}_{l} (\mu) \,  := \, \sum_{k \in \Lambda^{*}_{l}} (\varepsilon_{k} - \mu) a^{*}_{k}a_{k} \ ,
\end{eqnarray}
with quadratic spectrum and $\varepsilon_{k=0} =0$.

It is well known that this system exhibits a generalised condensation of type II for $\alpha_x = 1/2$ and
of type III for $\alpha_x > 1/2$ for a standard critical density $\rho_{c}$, whereas for $\alpha_x < 1/2$,
the whole condensate is sitting in the mode $k=0$ , i.e, in the ground state (type I) \cite{vdBLP}.
Consider the system (\ref{freeGBEC}) with the quasi-average source in a single mode $\tilde{k}$:
\begin{eqnarray}\label{freeQE}
H^{0}_{l} (\mu; \eta) \, := \, H^{0}_{l} (\mu) \, + \, \sqrt{V_{l}} \ \big( \overline{\eta} \  a_{\tilde{k}} +
\eta \ a^{*}_{\tilde{k}} \big) \ .
\end{eqnarray}
Then for a fixed density $\overline{\rho}$, the finite-volume equation which defines the corresponding chemical
potential $\mu = \overline{\mu}_{l} (\overline{\rho}, \eta)$ takes the form:
\begin{eqnarray}\label{perfect-gas-with-source-density-equation-finite-volume}
\overline{\rho} \, &=& \, \rho_{l}(\beta, \mu, \eta) \, := \, \frac{1}{V_{l}} \sum_{k \in \Lambda^{*}_{l}}
\langle a^{*}_{k}a_{k}\rangle^{0}_{l}(\beta,\mu,\eta) \\
&=& \, \frac{1}{V_{l}} (e^{\beta(\varepsilon_{\tilde{k}} - \mu)}-1)^{-1} \, + \, \frac{1}{V_{l}}
\sum_{k \neq \tilde{k}} \frac{1}{e^{\beta(\varepsilon_{k} - \mu)}-1} \, + \, \frac{\vert \eta \vert\, ^{2}}
{(\varepsilon_{\tilde{k}} - \mu)\, ^{2}} \ . \nonumber
\end{eqnarray}
To investigate a possible condensation, we must take the thermodynamic limit in the right-hand side of
(\ref{perfect-gas-with-source-density-equation-finite-volume}), and then switch off the source,
that is let $\vert \eta \vert \rightarrow 0$.
Let us denote by $I(\beta, \mu)$ the limit of $\rho_{l}(\beta, \mu, \eta = 0)$, that is the limiting density function
of the \emph{gauge-invariant} system,
\begin{eqnarray*}
I(\beta, \mu) \, = \, \lim_{l \rightarrow \infty} \, \frac{1}{V_{l}}
\sum_{k \neq \tilde{k}} \frac{1}{e^{\beta(\varepsilon_{k} - \mu)}-1} \, = \,
\int_{\mathbb{R}} \nu^{0} (\ud \varepsilon) \frac{1}{e^{\beta(\varepsilon - \mu)}-1} \ .
\end{eqnarray*}
with critical density $\rho_{c}:= \sup_{\mu<0} I(\mu)$.

Now we have to distinguish two cases:\\
(i) For any $\tilde{k}$ such that $\lim_{l \rightarrow \infty} \varepsilon_{\tilde{k}} > 0$, we obtain from
(\ref{perfect-gas-with-source-density-equation-finite-volume})
\begin{eqnarray*}
\overline{\rho} \, = \, \lim_{\vert \eta \vert \rightarrow 0} 
\lim_{l \rightarrow \infty} \rho_{l}(\beta, \mu, \eta) \, = \, I(\beta, \mu)
\end{eqnarray*}
i.e. the quasi-average coincides with the average and we return to the analysis of the condensate equation
(\ref{perfect-gas-with-source-density-equation-finite-volume}) for $\eta =0$. This gives again all
possible types of condensation as a function of $\alpha_x$.\\
(ii) On the other hand, if $\tilde{k}$ is such that $\lim_{l \rightarrow \infty} \varepsilon_{\tilde{k}} = 0$,
then the condensate equation
(\ref{perfect-gas-with-source-density-equation-finite-volume}) yields for the quasi-average
of the total particle density:
\begin{eqnarray}\label{perfect-gas-with-source-density-equation-infinite-volume}
\overline{\rho} \, = \,  \lim_{\vert \eta \vert \rightarrow 0} \lim_{l \rightarrow \infty} \rho_{l}(\beta, \mu, \eta)
\, = \, I(\beta, \mu) + \lim_{\eta \rightarrow 0} \frac{\vert \eta \vert\, ^{2}}{\mu\, ^{2}} \ .
\end{eqnarray}
If $\overline{\rho} \leq \rho_{c}$, then the asymptotic solution of
(\ref{perfect-gas-with-source-density-equation-infinite-volume}) is $\overline{\mu}_{\infty} (\overline{\rho}) =
\lim_{\eta \rightarrow 0}\lim_{l \rightarrow \infty}\overline{\mu}_{l} (\overline{\rho}, \eta) <0$ and
there is no condensation in any mode.\\
If $\overline{\rho} > \rho_{c}$, then $\lim_{\eta \rightarrow 0}
{\vert \eta \vert\, ^{2}}/{\overline{\mu}_{\infty} (\overline{\rho},\eta)\, ^{2}} = \overline{\rho} - \rho_{c}$.
By explicit calculation one also gets that only $\tilde{k}$-mode quasi-average is non-zero:
\begin{equation}\label{BEC-onemode}
\lim_{\eta \rightarrow 0}\lim_{l \rightarrow \infty} \frac{1}{V_l} \ 
\langle a^{*}_{\tilde{k}}a_{\tilde{k}}\rangle^{0}_{l}(\beta,\overline{\mu}_{l},\eta)=
\lim_{\eta \rightarrow 0}\lim_{l \rightarrow \infty} 
\left\{\frac{1}{V_l} \, \frac{1}{e^{\beta(\varepsilon_{\tilde{k}}- \overline{\mu}_{l} (\overline{\rho}, \eta))}-1} 
+ \frac{\vert \eta \vert\, ^{2}}{\overline{\mu}_{l}\, ^{2}}\right\} = \overline{\rho} - \rho_{c} \ ,
\end{equation}
i.e. for any $\alpha_x$ the condensation is type I. Recall that the only condition on $\tilde{k}$ is that the 
corresponding eigenvalue
$\varepsilon_{\tilde{k}}$ vanishes in the infinite volume limit.
Since
\begin{equation}\label{zero-non-zero-modes}
\lim_{\eta \rightarrow 0}\lim_{l \rightarrow \infty} \frac{1}{V_l} \ 
\langle a^{*}_{0}a_{0}\rangle^{0}_{l}(\beta,\overline{\mu}_{l},\eta) =
\lim_{\eta \rightarrow 0}\lim_{l \rightarrow \infty} \frac{1}{V_l} 
\frac{1}{e^{\beta(- \overline{\mu}_{l} (\overline{\rho}, \eta))}-1} = 0 \ ,
\end{equation}
by virtue of  (\ref{BEC-onemode}) and (\ref{zero-non-zero-modes}) we see that the Bogoliubov quasi-average procedure 
not only transforms the generalised condensates of type II or III into a one-mode condensate (i.e., type I), 
but this mode does not even need to be the ground-state. Therefore, using the quasi-average approach \cite{Bog}, one 
can force the condensate to be in any given mode $\tilde{k}$, as long as its energy $\varepsilon_{\tilde{k}}$ 
vanishes in the limit $l \rightarrow \infty$.

\section*{Appendix}
\appendix
\section{The approximating Hamiltonians}\label{Appendix-explicit-expression-lower-upper-symbol}
\renewcommand{\theequation}{\Alph{section}.\arabic{equation}}
\renewcommand{\thelemma}{\Alph{section}.\arabic{lemma}}
\setcounter{section}{1}
\setcounter{equation}{0}
\setcounter{theorem}{0}
\setcounter{lemma}{0}

In this section, we provide an explicit form of the upper and lower symbols for the Hamiltonian
(\ref{multi-part-interacting-Hamiltonian}). For a short-hand we put $E_{i}^{\omega,l} = E_{i}$.
We note that, as the coherent vector (\ref{coherent-vector-definition})
is defined as a tensor product
of one-mode coherent states, its effect on each creation/annihilation operator $a^{\sharp}_{k}$ is independent
of all the others modes operators. First, we give an explicit form of the lower symbol of the full Hamiltonian 
(\ref{multi-part-interacting-Hamiltonian}):
{\allowdisplaybreaks
\begin{eqnarray}
H_{l}^{low}(\mu, \{c_{k}\})&=&  \sum_{k \in I_{\delta}} \big( \sum_{i \geqslant 1} |(\phi_{i},\psi_{k})|\, ^{2}
(E_{i} - \mu) \big) \vert c_{k}\vert\, ^{2}
\label{full-hamiltonian-one-particle-energy-term1}\\
&+& \sum_{k \in I_{\delta}^{c}} \big( \sum_{i \geqslant 1} |(\phi_{i},\psi_{k})|\, ^{2} 
(E_{i} - \mu) \big) a^{*}_{k}a_{k}
\label{full-hamiltonian-one-particle-energy-term2}\\
&+&  \sum_{k\in I_{\delta}, k' \in I_{\delta}^{c}} \big( \sum_{i \geqslant 1} 
(\phi_{i},\psi_{k})(\psi_{k'},\phi_{i})
(E_{i} - \mu) \big) \overline{c_{k}} a_{k'}\label{full-hamiltonian-one-particle-energy-term3}\\
&+&  \sum_{k\in I_{\delta}^{c}, k' \in I_{\delta}} \big( \sum_{i \geqslant 1} (\phi_{i},\psi_{k})(\psi_{k'},\phi_{i})
(E_{i} - \mu) \big) c_{k} a^{*}_{k}\label{full-hamiltonian-one-particle-energy-term4}\\
&+&  \sum_{k, k' \in I_{\delta}^{c}, k \neq k'} \big( \sum_{i \geqslant 1} (\phi_{i},\psi_{k})(\psi_{k'},\phi_{i})
(E_{i} - \mu) \big) a^{*}_{k}a_{k'}\label{full-hamiltonian-one-particle-energy-term5}\\
&+& \frac{1}{2V_{l}} \sum_{k\in I_{\delta}} \, \sum_{k'\in I_{\delta}} \,\, \sum_{\substack{q:\, 
k+q\in I_{\delta}^{c} \\
k'-q\in I_{\delta}^{c}}} \hat{u}(q) \, \, c_{k} c_{k'} \, a^{*}_{k+q} a^{*}_{k'-q}
\label{full-hamiltonian-interaction-term1}\\
&+& \frac{1}{2V_{l}} \sum_{k\in I_{\delta}} \, \sum_{k'\in I_{\delta}} \,\, \sum_{\substack{q:\, k+q\in I_{\delta} \\
k'-q\in I_{\delta}^{c}}} \hat{u}(q)  \, \, \overline{c}_{k+q} c_{k} c_{k'} \, a^{*}_{k'-q}
\label{full-hamiltonian-interaction-term2}\\
&+& \frac{1}{2V_{l}} \sum_{k\in I_{\delta}} \, \sum_{k'\in I_{\delta}} \,\, \sum_{\substack{q:\, 
k+q\in I_{\delta}^{c} \\
k'-q\in I_{\delta}}} \hat{u}(q)  \, \,  \overline{c}_{k'-q} c_{k} c_{k'} \, a^{*}_{k+q}
\label{full-hamiltonian-interaction-term3}\\
&+& \frac{1}{2V_{l}} \sum_{k\in I_{\delta}} \, \sum_{k'\in I_{\delta}} \,\, \sum_{\substack{q:\, k+q\in I_{\delta} \\
k'-q\in I_{\delta}}} \hat{u}(q)  \, \,  \overline{c}_{k+q}  \overline{c}_{k'-q}c_{k} c_{k'}
\label{full-hamiltonian-interaction-term4}\\
&+& \frac{1}{2V_{l}} \sum_{k\in I_{\delta}} \, \sum_{k'\in I_{\delta}^{c}} \,\, \sum_{\substack{q:\,
k+q\in I_{\delta}^{c} \\
k'-q\in I_{\delta}^{c}}} \hat{u}(q)  \, \, c_{k'} \, a^{*}_{k+q} a^{*}_{k'-q} a_{k}
\label{full-hamiltonian-interaction-term5}\\
&+& \frac{1}{2V_{l}} \sum_{k\in I_{\delta}} \, \sum_{k'\in I_{\delta}^{c}} \,\, 
\sum_{\substack{q:\, k+q\in I_{\delta} \\
k'-q\in I_{\delta}^{c}}} \hat{u}(q)  \, \, \overline{c}_{k+q} c_{k} \, a^{*}_{k'-q} a_{k'}
\label{full-hamiltonian-interaction-term6}\\
&+& \frac{1}{2V_{l}} \sum_{k\in I_{\delta}} \, \sum_{k'\in I_{\delta}^{c}} \,\, \sum_{\substack{q:\,
k+q\in I_{\delta}^{c} \\
 k'-q\in I_{\delta}}} \hat{u}(q)  \, \, \overline{c}_{k'-q} c_{k} \, a^{*}_{k+q} a_{k'}
 \label{full-hamiltonian-interaction-term7}\\
&+& \frac{1}{2V_{l}} \sum_{k\in I_{\delta}} \, \sum_{k'\in I_{\delta}^{c}} \,\, 
\sum_{\substack{q:\, k+q\in I_{\delta} \\
k'-q\in I_{\delta}}} \hat{u}(q)  \, \, \overline{c}_{k+q} \overline{c}_{k'-q}c_{k} \, a_{k'}
\label{full-hamiltonian-interaction-term8}\\
&+& \frac{1}{2V_{l}} \sum_{k\in I_{\delta}^{c}} \, \sum_{k'\in I_{\delta}} \,\,
\sum_{\substack{q:\, k+q\in I_{\delta}^{c} \\
k'-q\in I_{\delta}^{c}}} \hat{u}(q)  \, \, c_{k'} \, a^{*}_{k+q} a^{*}_{k'-q} a_{k}
\label{full-hamiltonian-interaction-term9}\\
&+& \frac{1}{2V_{l}} \sum_{k\in I_{\delta}^{c}} \, \sum_{k'\in I_{\delta}} \,\, 
\sum_{\substack{q:\, k+q\in I_{\delta} \\
k'-q\in I_{\delta}^{c}}} \hat{u}(q)  \, \, \overline{c}_{k+q} c_{k'} \, a^{*}_{k'-q} a_{k}
\label{full-hamiltonian-interaction-term10}\\
&+& \frac{1}{2V_{l}} \sum_{k\in I_{\delta}^{c}} \, \sum_{k'\in I_{\delta}} \,\,
\sum_{\substack{q:\, k+q\in I_{\delta}^{c} \\
 k'-q\in I_{\delta}}} \hat{u}(q)  \, \, \overline{c}_{k'-q} c_{k'} \, a^{*}_{k+q} a_{k}
 \label{full-hamiltonian-interaction-term11}\\
&+& \frac{1}{2V_{l}} \sum_{k\in I_{\delta}^{c}} \, \sum_{k'\in I_{\delta}} \,\, 
\sum_{\substack{q:\, k+q\in I_{\delta} \\
 k'-q\in I_{\delta}}} \hat{u}(q)  \, \, \overline{c}_{k+q} \overline{c}_{k'-q} c_{k'} \, a_{k}
 \label{full-hamiltonian-interaction-term12}\\
&+& \frac{1}{2V_{l}} \sum_{k\in I_{\delta}^{c}} \, \sum_{k'\in I_{\delta}^{c}} \,\,
\sum_{\substack{q:\, k+q\in I_{\delta}^{c} \\
 k'-q\in I_{\delta}^{c}}} \hat{u}(q)  \, \, a^{*}_{k+q} a^{*}_{k'-q} a_{k'} a_{k}
 \label{full-hamiltonian-interaction-term13}\\
&+& \frac{1}{2V_{l}} \sum_{k\in I_{\delta}^{c}} \, \sum_{k'\in I_{\delta}^{c}} \,\,
\sum_{\substack{q:\, k+q\in I_{\delta} \\
 k'-q\in I_{\delta}^{c}}} \hat{u}(q) \, \, \overline{c}_{k+q} \,  a^{*}_{k'-q} a_{k'} a_{k}
 \label{full-hamiltonian-interaction-term14}\\
&+& \frac{1}{2V_{l}} \sum_{k\in I_{\delta}^{c}} \, \sum_{k'\in I_{\delta}^{c}} \,\,
\sum_{\substack{q:\, k+q\in I_{\delta}^{c} \\
 k'-q\in I_{\delta}}} \hat{u}(q) \, \, \overline{c}_{k'-q} \, a^{*}_{k+q}  a_{k'} a_{k}
 \label{full-hamiltonian-interaction-term15}\\
&+& \frac{1}{2V_{l}} \sum_{k\in I_{\delta}^{c}} \, \sum_{k'\in I_{\delta}^{c}} \,\,
\sum_{\substack{q:\, k+q\in I_{\delta} \\
 k'-q\in I_{\delta}}} \hat{u}(q)  \, \, \overline{c}_{k+q} \overline{c}_{k'-q} \, a_{k'} a_{k}
 \label{full-hamiltonian-interaction-term16}
\end{eqnarray}}

Now, we give an explicit form of the upper symbols. We recall the general form of this symbols for polynomials in the
creation/annihilation operators of some mode $k \in I_{\delta}$:
\begin{eqnarray*}
(a_{k})^{up} = c_{k}, \quad (a_{k}^{*})^{up} = \overline{c}_{k}, \quad (a_{k} a_{k})^{up} =
(c_{k})\, ^{2}, \quad (a_{k}^{*} a_{k})^{up} = (\overline{c}_{k})\, ^{2}\\
\quad (a_{k}^{*} a_{k})^{up} = \vert c_{k}\vert\, ^{2} - 1, \quad \quad (a_{k}^{*}a_{k}^{*} a_{k}a_{k})^{up} =
\vert c_{k}\vert^{4} - 4\vert c_{k}\vert\, ^{2} + 2
\end{eqnarray*}
Note that, since the interaction term of the Hamiltonian term considered on its own does
have a momentum conservation law, it is not possible to get exactly three out of four operators in the same mode $k$.
In view of this, it can be seen that the lower and upper symbols of the Hamiltonian will differ only when two or four
operators in the same mode appears, that is only terms (\ref{full-hamiltonian-one-particle-energy-term1}),
(\ref{full-hamiltonian-interaction-term4}), (\ref{full-hamiltonian-interaction-term7}),
(\ref{full-hamiltonian-interaction-term8}), (\ref{full-hamiltonian-interaction-term10}),
(\ref{full-hamiltonian-interaction-term11}) and (\ref{full-hamiltonian-interaction-term12})
differs in both approximating Hamiltonians.\\
Splitting further the sums in these terms leads finally to the final upper symbol of the Hamiltonian
\begin{eqnarray*}
H_{l}^{\textrm{up}} (\mu, \{c_{k}\}) \, = \, H_{l}^{\textrm{low}} (\mu, \{c_{k}\}) + \kappa(\mu, \{c_{k}\}),
\end{eqnarray*}
where
{\allowdisplaybreaks
\begin{eqnarray}\label{appendix-Bogoliubov-expression-kappa}
\kappa(\mu, \{c_{k}\})  &=& \sum_{k \in I_{\delta}} \big( \sum_{i \leqslant 1} |(\phi_{i},\psi_{k})|\, ^{2}
(E_{i} - \mu) \big)\\ \nonumber
&+& - \frac{1}{2V_{l}} \Big( 2 \sum_{k \in I_{\delta}} \hat{u}(0) +  \sum_{\substack{k,k' \in I_{\delta}
\\ k \neq k'}} \hat{u}(0) +
\sum_{q \neq 0} \sum_{k \in I_{\delta} \cap I-q} \hat{u}(q)\Big)\\
&+& \frac{1}{2V_{l}} \Big( 4 \sum_{k \in I_{\delta}} \hat{u}(0) \vert c_{k}\vert\, ^{2} + \sum_{\substack{k,k'
\in I_{\delta} \\ \nonumber
k \neq k'}} \hat{u}(0)
(\vert c_{k}\vert\, ^{2} + \vert c_{k'}\vert\, ^{2}) +  \sum_{q \neq 0} \sum_{\substack{k \in I_{\delta} \cap I-q
\\ \nonumber
k' \in I_{\delta} \cap I-q \\ k' - k = q}} \hat{u}(q) (\vert c_{k}\vert\, ^{2} + \vert c_{k'}\vert\, ^{2})
\Big)\\ \nonumber
&+&  \frac{1}{2V_{l}} \Big( 2 \sum_{k \in I_{\delta}} \hat{u}(0) \sum_{k' \in  I_{\delta}^{c}} a^{*}_{k'} a_{k'} +
\sum_{q \neq 0} \sum_{\substack{k \in I_{\delta} \\ \nonumber
k' \in I_{\delta}^{c} \cap I+q \\ k' - k = q}} \hat{u}(q) a^{*}_{k} a_{k} +
\sum_{q \neq 0} \sum_{\substack{k \in I_{\delta}^{c} \cap I-q \\ k' \in I_{\delta} \\ k' - k = q}}
\hat{u}(q) a^{*}_{k} a_{k}\Big) \ .
\end{eqnarray}
}
\section*{Acknowledgments}

One of the authors (Th.Jaeck) is supported by funding from the UCD Ad Astra Research Scholarship.
We are thankful to J.V.Pul\'{e} and J-B Bru for very useful remarks and suggestions.


\begin{thebibliography}{999}

\bibitem{B} N.N. Bogoliubov, \textit{Izv. Akad. Nauk USSR} \textbf{11} (1947) 77-90
\bibitem{Bog} N.N. Bogoliubov, \textit{Lectures on Quantum Statistics}, Gordon and Breach, Paris 1968 ;
Selected Papers, vol.VIII, Moscow 2009
\bibitem{FPV} M. Fannes, J.V. Pulé, A. Verbeure, \textit{Helv.Phys.Acta} \textbf{55} (1982), 391-399
\bibitem{G} J. Ginibre, \textit{Commun. Math. Phys.} \textbf{8} (1968), 26-51
\bibitem{LSY} E.H. Lieb, R. Seiringer, J. Yngvason,  \textit{Phys. Rev. Lett.} \textbf{94} (2005), 080401
\bibitem {vdBLP} M. van den Berg, J.T.Lewis, J.V.Pul\'{e}, \textit{Helv.Phys.Acta} \textbf{59} (1986), 1271-1288
\bibitem{BdSP} E. Buffet, Ph. de Smedt, J.V. Pulé, J. Phys. A: Math. Gen. \textbf{16} (1983), 4307-4324
\bibitem{LGP} I.M.Lifshitz, S.A.Gredeskul, L.A.Pastur, \textit{Introduction to the Theory of Disordered\\ Systems},
Wiley, N.Y. 1989
\bibitem {KL1} M. Kac, J.M. Luttinger, \textit{J. Math. Phys.} \textbf{14} (1973), 1626-1628
\bibitem {KL2} M. Kac, J.M. Luttinger, \textit{J. Math. Phys.} \textbf{15} (1974), 183-186
\bibitem {LS} J.M. Luttinger, H.K. Sy, \textit{Phys. Rev. A} \textbf{7} (1973), 712-720
\bibitem {LPZ} O. Lenoble, L.A. Pastur, V.A. Zagrebnov, \textit{Comptes-rendus de
l'Acad\'{e}mie des Sciences (Paris), Physique} \textbf{5} (2004), 129-142
\bibitem {LZ} O. Lenoble, V.A. Zagrebnov, \textit{Markov Processes and related fields} \textbf{13} (2007), 441-468
\bibitem {ZB} V.A. Zagrebnov, J.-B. Bru, \textit{Phys. Rep.} \textbf{350} (2001), 291-434
\bibitem{JPZ1} T. Jaeck, J.V. Pul\'e, V.A. Zagrebnov,  \textit{J. Stat. Phys.}  \textbf{137}  (2009), 19-55
\bibitem{JPZ2} T. Jaeck, J.V. Pul\'e, V.A. Zagrebnov, \textit{J. Math. Phys.}  \textbf{XX}  (2010), YY-ZZ
\bibitem {HM} K. Huang, H.F. Men, \textit{Phys. Rev. Lett.} \textbf{69} (1992),  644-647
\bibitem {KT} M. Kobayashi, M. Tsubota,\textit{ Phys. Rev.} \textbf{66} (2002), 174516
\bibitem{Rob} D.W. Robinson, \textit{The thermodynamic pressure in quantum statistical mechanics},
Lecture Notes in Physics, Vol. 9. Springer-Verlag, Berlin-New York 1971
\bibitem{AN} N. Angelescu, G. Nenciu, \textit{Commun. Math. Phys.} \textbf{29} (1973), 15-30
\bibitem{Ste} J.M. Steele \textit{Ann. Inst. H. Poincaré Probab. Statist.}  \textbf{25} (1989), no.1, 93-98
\bibitem{PZ} J.V.Pul\'{e}, V.A.Zagrebnov, \textit{ Reviews in Math.Phys.} \textbf{19} (2007), 157-194
\bibitem{BZ} J.-B.Bru, V.A.Zagrebnov, \textit{J.Stat.Phys.} {\bf 133} (2008), 379-400


\end{thebibliography}
\end{document}